\documentclass[sn-mathphys]{sn-jnl}% Math and Physical Sciences Reference Style
\usepackage{amsmath}
\jyear{2021}%

\theoremstyle{thmstyleone}%
\theoremstyle{thmstyletwo}%

\theoremstyle{thmstylethree}%

\begin{document}

\title[Main text]{Simultaneously Enhanced Tenacity, Rupture Work, and Thermal Conductivity of Carbon Nanotubes Fibers by Increasing the Effective Tube Contribution}


\author[1,2]{\fnm{Xiao} \sur{Zhang}}\email{zhangx@iphy.ac.cn}

\author*[2]{\fnm{Michael} \sur{De Volder}}\email{mfld2@cam.ac.uk}

\author[3]{\fnm{Wenbin} \sur{Zhou}}\email{wbzhou@bjut.edu.cn}

\author[2]{\fnm{Liron} \sur{Issman}}\email{li242@cam.ac.uk}

\author[1]{\fnm{Xiaojun} \sur{Wei}}\email{weixiaojun@iphy.ac.cn}

\author[4]{\fnm{Adarsh} \sur{Kaniyoor}}\email{ak2011@cantab.ac.uk}

\author[4]{\fnm{Jerónimo Portas} \sur{Terrones}}\email{jt451@cam.ac.uk}

\author[2]{\fnm{Fiona} \sur{Smail}}\email{frs25@cam.ac.uk}

\author[1]{\fnm{Zibo} \sur{Wang}}\email{wangzibo@iphy.ac.cn}

\author[1]{\fnm{Yanchun} \sur{Wang}}\email{ycwang@iphy.ac.cn}

\author[1]{\fnm{Huaping} \sur{Liu}}\email{liuhuaping@iphy.ac.cn}

\author[1]{\fnm{Weiya} \sur{Zhou}}\email{wyzhou@iphy.ac.cn}

\author*[4]{\fnm{James} \sur{Elliott}}\email{jae1001@cam.ac.uk}

\author*[1]{\fnm{Sishen} \sur{Xie}}\email{ssxie@iphy.ac.cn}

\author*[2]{\fnm{Adam} \sur{Boies}}\email{amb233@cam.ac.uk}

\affil[1]{\orgdiv{Beijing National Laboratory for Condensed Matter Physics}, \orgname{Institute of Physics, Chinese Academy of Sciences}, \city{Beijing}, \postcode{100190}, \country{China}}

\affil[2]{\orgdiv{Department of Engineering}, \orgname{University of Cambridge}, \city{Cambridge}, \postcode{CB2 1PZ}, \country{UK}}

\affil[3]{\orgdiv{MOE Key Laboratory of Enhanced Heat Transfer and Energy Conservation, Beijing Key Laboratory of Heat Transfer and Energy Conversion}, \orgname{Beijing University of Technology}, \city{Beijing}, \postcode{100124}, \country{China}}

\affil[4]{\orgdiv{Department of Materials Science and Metallurgy}, \orgname{University of Cambridge}, \city{Cambridge}, \postcode{CB3 0FS}, \country{UK}}

\abstract{Although individual carbon nanotubes (CNTs) are superior as constituents to polymer chains, the mechanical and thermal properties of CNT fibers (CNTFs) remain inferior to commercial synthetic fibers due to the lack of synthesis methods to embed CNTs effectively in superstructures. The application of conventional techniques for mechanical enhancement resulted in a mild improvement of target properties while achieving parity at best on others. In this work, a Double-Drawing technique is developed to deform continuously grown CNTFs and rearrange the constituent CNTs in both mesoscale and nanoscale morphology. Consequently, the mechanical and thermal properties of the resulting CNTFs can be jointly improved, and simultaneously reach their highest performances with specific strength (tenacity) $\rm\sim3.30\,N\,tex^{-1}$, work of rupture $\rm\sim70\,J\,g^{-1}$, and thermal conductivity $\rm\sim354\,W\,m^{-1}\,K^{-1}$, despite starting from commercial low-crystallinity materials ($I{\rm_G}:I{\rm_D}\sim5$). The processed CNTFs are more versatile than comparable carbon fiber, Zylon, Dyneema, and Kevlar. Furthermore, based on evidence of load transfer efficiency on individual CNTs measured with In-Situ Stretching Raman, we find the main contributors to property enhancements are (1) the increased proportion of load-bearing CNT bundles and (2) the extension of effective length of tubes attached on these bundles.}




\maketitle
\section{Introduction}\label{Introduction}

Carbon nanotube (CNT) macroscopic assemblies, like CNT fibers (CNTFs) are analogous to bulk materials of highly-conjugated polymer molecules.\cite{Fakhri2009,Mikhalchan2019} Correspondingly, constituent CNTs are akin to polymer chains, but with outstanding mechanical,\cite{Min-Feng2000,Peng2008,Zhang2020d} thermal,\cite{Kim2001,Pop2006} electrical properties,\cite{Bulmer2021,Komatsu2021} and chemical resilience.\cite{DeVolder2013} CNT bundles are reported to possess excellent tensile strength ($\rm27\!-\!31\,N\,tex^{-1}$) and Young's modulus ($\rm337\!-\!640\,N\,tex^{-1}$).\cite{Tombler2000,Yu2000,Baughman2002,Bai2018} Direct-spun CNTFs can be produced continuously ($\rm1\!-\!2\,km\,h^{-1}$) with low-cost, resulting in light-weight and high flexibility fibers. However, raw direct-spun CNTFs still suffer from a 1-2 orders of magnitude degradation in properties relative to CNT bundles, and are uncompetitive with commercial carbon fibers (CF), Kevlar fibers and Zylon fibers on their corresponding specialties.\par
Researchers attribute the property degradation primarily to the poor arrangement of the constituent CNTs within the as-synthesized CNTFs,\cite{Mikhalchan2019} and thus they utilized various post-synthesis treatments seeking to rearrange the individual CNTs (iCNTs) and CNT bundles. However, with techniques commonly used on textile fibers (e.g. direct stretching,\cite{Koziol2007,Fernandez-Toribio2018} compression,\cite{Wang2014a,Tran2016,Xu2016} and twisting\cite{Zhang2004,Headrick2018}) only minor enhancements have been achieved. In 2013, with a similar method to produce Kevlar and Zylon, the solution-spinning technique was reported to produce highly aligned and compacted CNTFs\cite{Behabtu2013} from a liquid crystal solution of CNTs in chlorosulfonic acid (CSA). The reported mechanical properties increased significantly to strengths $\rm\sim1.0\,GPa$ (tenacity $\rm\sim0.97\,N\,tex^{-1}$) and moduli $\rm\sim120\,GPa$. Recently researchers stretched the Direct-spun CNTFs in CSA to obtain an aligned structure\cite{Lee2019} which employed screening of van der Waals (vdW) forces in CSA,\cite{Davis2009,Parra-Vasquez2010,Ramesh2004} similar to stiffening cellulose fibers within water to screen hydrogen bonds between chains.\cite{Hearle2008} Even though the reported mechanical performances surpass those of solution-spinning CNTFs, superior performance seems only achievable on very thin CNTFs (linear density $\rm0.05\,tex$), while those of thick CNTFs are much degraded, even poorer than raw thin CNTFs.\cite{Lee2020} The achieved density is limited to less than $\rm1.1\,g\,cm^{-3}$, indicating potential for further enhancement to the theoretical density of ideally compacted CNTs bundles ($\rm1.54\,g\,cm^{-3}$).\cite{Aliev2010,Behabtu2013} Meanwhile, CSA appears to have competing effects, facilitating rearrangement while hindering the load transfer within CNTFs in internal regions where remnant CSA screens vdW forces.\cite{Ramesh2004} Additionally, because phonon conduction is also impeded by the disordered microstructure in CNTFs, strategies to enhance thermal conductivity also warrant investigation. Thus, improving the properties of CNTFs while avoiding the disturbance from residual CSA is the primary motivation of our work.\par
Additionally, while many improvements have been reported for CNT performance, a mechanistic explanation of property enhancement is still incomplete. Most conceptual models omit the distinctive structure of CNTFs as a porous hierarchical network of long rigid iCNTs, as distinct from CF, Kevlar, or cotton yarn. Furthermore, the efficiency of load transfer between iCNTs, and focus on the different behavior of iCNTs under various loads remains to be fully investigated. Thus, our work also focuses on optimizing the mechanism applicable to hierarchical CNT networks bonded by vdW forces to guide further enhancement and to assess current processing techniques.\par
Here, we develop a novel Double-Drawing technique to straighten and compact the iCNTs within raw Direct-spun CNTFs, seeking a simultaneous improvement in mechanical and thermal properties. The iCNT alignment and CNTF porosity were monitored by Wide and Small-Angle X-ray Scattering and FIB cross-section analysis, as well as in-situ stretching Raman to study quantitatively the iCNTs' behavior in various CNTFs after different levels of processing and loading. Finally, based on the experimental findings, we complement the mechanism of enhancement with two critical factors: the increased proportion of load-bearing CNT bundles and the extension of effective length of tubes attached on these bundles.\par

\section{Results}\label{Results}
\subsection{Enhancement of CNTF with the Double-Drawing Process}\label{Enhancement of CNTF with the Double-Drawing Process}
\begin{figure}[H]%
    \centering
    \includegraphics[width=0.9\textwidth]{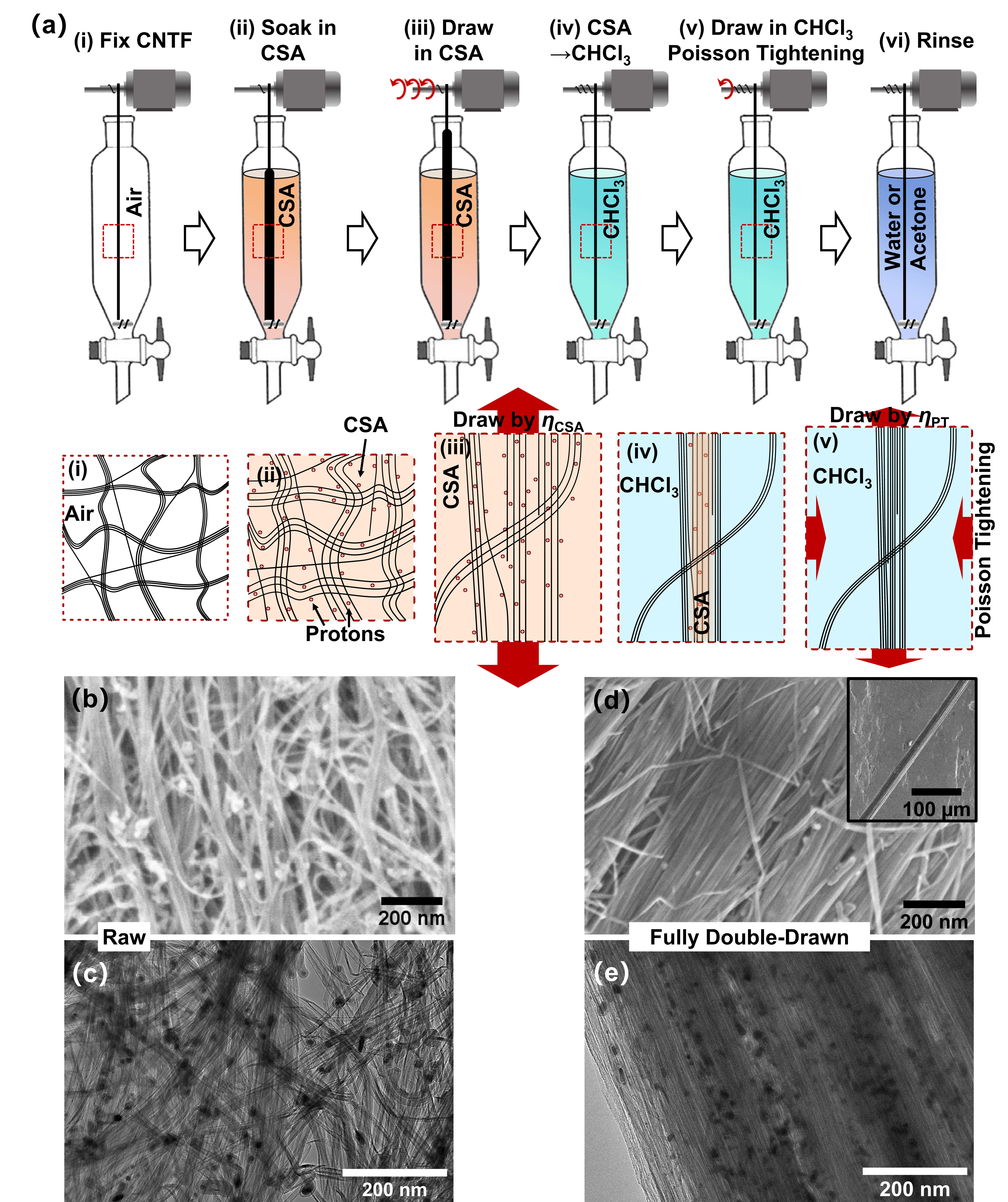}
    \caption{The Double-Drawing process on CNT fiber (CNTF). (a) Schematic diagram of the Double-Drawing process to enhance the CNTFs, and the corresponding typical cell within CNTF (second row): (i) the raw CNTF is fixed within a dropping funnel straightened out but without a pre-tension; (ii) to weaken the binding between iCNTs, the raw CNTF is immersed in HSO3Cl (CSA); (iii) CNTF is then firstly drawn in CSA to a specific ratio ($\eta{\rm_{CSA}}$) to straighten out and align the crumpled iCNTs; (iv) after the immersing solvent being changed into CHCl3 (chloroform), (v) CNTF is further drawn by $\eta{\rm_{PT}} = 0.5\%$, referred as Poisson Tightening. During this process, the remaining CSA is removed from the fiber, which radially compresses and ``freezes" iCNTs into larger bundles; (vi) Finally CNTF is successively rinsed in water and acetone, and vacuum dried. The substantial enhancement on alignment and compactness of fully Double-Drawn CNTF can be observed on the surface and inner microstructures by (b vs d) SEM and (c vs e) HRTEM. The red arrows in the schematic cell (the second row) represent the force exerted.}\label{Fig:1}
\end{figure}

In our work, the raw CNTFs are fabricated by direct spinning CNT aerogels produced with a continuous floating catalyst chemical vapor deposition (FCCVD) method. The FCCVD method is considered highly suitable for the continuous mass-production of iCNTs with a very high aspect ratio.\cite{Mikhalchan2019} These grown CNTs aggregate into bundles and then entangle as an aerogel,\cite{Boies2019,Kateris2020} which is subsequently densified into fibers by acetone during collection from the reactor. The as-synthesized CNTFs consist of a hierarchical network of randomly connected CNTs (Fig.\ref{Fig:1}a(i), b-c). Analogous to polymer molecules in a textile fiber being held together by hydrogen bonds and/or vdW forces, the iCNTs in CNTF are held together by vdW forces in an entangled network. The vdW forces ``freeze" the CNTF in a non-equilibrium curved morphology, i.e. an athermal structure.\cite{Yakobson2006}\par
Here, to enhance the CNTF after synthesis, the connections or constraint intra and inter bundles are firstly weakened or even released by immersing the CNTF in $\rm HSO_3Cl$ (CSA) (Fig.\ref{Fig:1}a(ii)). The protonation of CNTs by CSA\cite{Davis2009} screens the vdW forces, as observed by the swelling of the CNTF in CSA. The weakened tube-tube forces thus reduce the large shear strength. Thus, the mutual lateral movement of iCNTs can occur easily, which otherwise might break CNTs if done without CSA. This process is analogous to the high humidity environment softening the hydrogen bonds between cellulose molecules before enhancing cellulose fiber.\par
The CNTF is then firstly drawn in CSA to a specific draw-ratio ($\eta{\rm_{CSA}}$, the extension length divided by original length), during which the crumpled CNTs are freely straightened and aligned along the fiber axis (Fig.\ref{Fig:1}a(iii)), with minimal breaking of CNTs. To quantitatively study the effect of drawing in CSA, the range of $\eta{\rm_{CSA}}$ is controlled from $0\%$ to the maximum ratio $\eta{\rm_{max}}$, in which the $\eta{\rm_{max}}\sim30\%$ less than the failure ratio in CSA ($\eta\rm^*$). We find a strong dependence of $\eta\rm^*$ on linear density ($LD$) of CNTF, i.e., $\eta\rm^*=41 \ \times \ $LD$^{1/2} [\%\,tex^{-1/2}]$ (details can be found in S1). For CNTFs the respective dependencies on $LD$ and draw-ratios are observed $LD\rm\sim0.5\,tex$, $\eta\rm^*\sim28\%$, and $\eta{\rm_{max}}\sim25\%$.\par
After drawing, the remaining CSA within CNTF still hinders the load transfer between iCNTs caused by vdW forces. Chloroform serves as the best solvent to dissolve CSA, but the fine voids in the increasingly compacted outer CNT layers hinder the outward diffusion of CSA (Fig.\ref{Fig:1}a(iv)). Therefore, to remove the remaining CSA and densify the fiber, we further introduce a second drawing process, referred as ``Poisson Tightening" process. After the first drawing in CSA, the fiber is immediately drawn further in chloroform with another draw-ratio, $\eta{\rm_{PT}}\sim0.5\%$ (Fig.\ref{Fig:1}a(v)). When immersed in chloroform, CNTs in the outer layer solidify, significantly, thus increasing the layer's modulus. Then under subsequent axial drawing, the resulting radial tightening, caused by the Poisson effect, further expels CSA, while solidify and compressing CNTs in the inner layers (details in S1). We also find excessive drawing, $\eta{\rm_{PT}}\textgreater0.5\%$, may cause plastic deformation and break iCNTs. After the above Double-Drawing processes, the CNTFs are successively rinsed with water and acetone (Fig.\ref{Fig:1}a(vi)), and finally vacuum dried for further use.\par

\subsection{The Microstructure After Drawing}\label{The Microstructure After Drawing}
\begin{figure}[H]%
    \centering
    \includegraphics[width=0.9\textwidth]{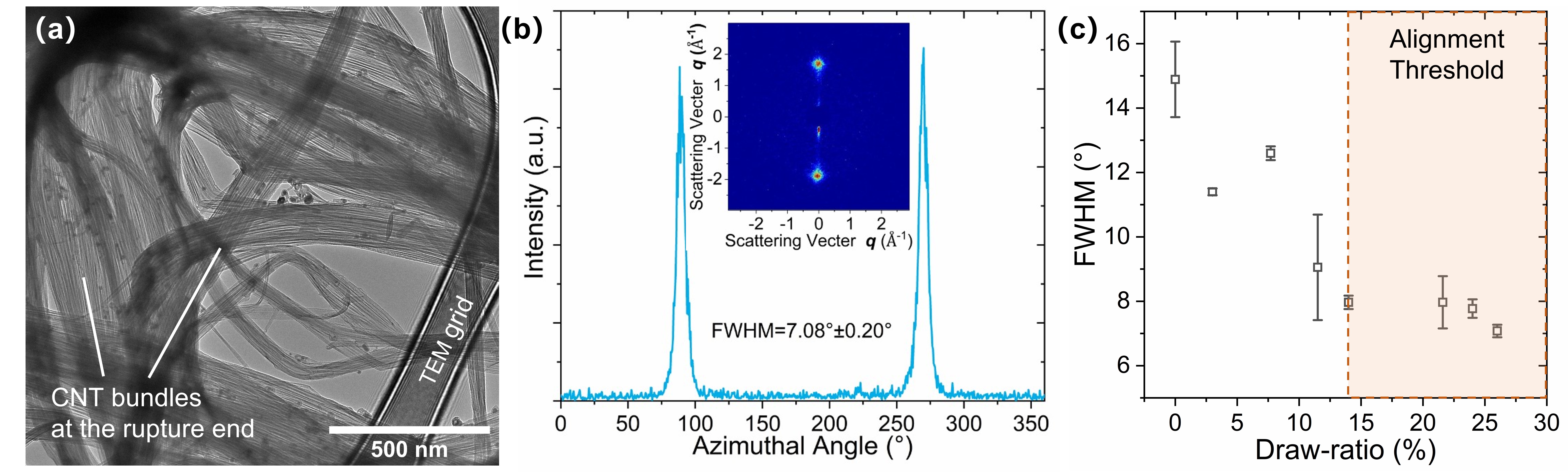}
    \caption{Microstructure of the CNTF after the Double-Drawing process. (a) The fracture end of the fully Double-Drawn CNTF illustrates fringed morphology with thick bundles commonly seen in the highly oriented linear-polymer fibers. Within every large bundle, the iCNTs closely compact into thick bundles in a range of 50-200 nm, much larger than those of raw CNTFs. (b) The Wide Angle X-ray Diffraction (WAXD) patterns (inset) of the suspended CNTFs illustrate the two preferred orientation peaks of the (002) planes in the azimuthal profile. The Full Width Half Maximum of the peaks indicate the level of alignment within CNTFs. (c) With the gradual increase in the draw-ratio, the alignment of iCNTs increases but reaches a plateau state after threshold level of drawing ($\rm\sim14.5\%$).}\label{Fig:2}
\end{figure}

For the Double-Drawn CNTF (DD-CNTF) with full drawing ($\eta{\rm_{CSA}}\!=\!\eta{\rm_{max}}\!=\!25\%$ and $\eta{\rm_{PT}}\!=\!0.5\%$), referred as fully DD-CNTF, both the mesoscale and nanoscale structure are organized. The disordered network of the raw CNTF is optimized into an aligned and tightly compacted bundle structure (Fig.\ref{Fig:1}b-e). The raw thin bundles (diameter of 10-50 nm, Fig.\ref{Fig:1}c) converge to much thicker bundles with diameter of 80-500 nm, as observed in the fracture end of the fully DD-CNTF (Fig.\ref{Fig:2}a). Within the thick bundles, all iCNTs are tightly packed.\par
The alignment evolution within CNTFs is observed by Wide-Angle X-ray Diffraction (WAXD).\cite{Bedewy2009,Fernandez-Toribio2018,Kaniyoor2021} The Full-Width-at-Half-Maximum (FWHM) of the preference peak in azimuthal scan is an indicator of CNTs alignment (Fig.\ref{Fig:2}b). As the draw-ratio increases to 14.5\%, the FWHM decreases from $\rm 14.89^\circ$ to $\rm 7.97^\circ$ (Fig.\ref{Fig:2}c). At draw-ratios above 14.5\%, FWHM plateaus near $\rm 7.9^\circ$, which indicates the saturation of the alignment of CNTs above the threshold level of drawing. Similar results are shown for Small-Angle X-ray Scattering (SAXS, Fig.S1) where the optimization of alignment levels off above 12\% of drawing.\par
\begin{figure}[H]%
    \centering
    \includegraphics[width=0.9\textwidth]{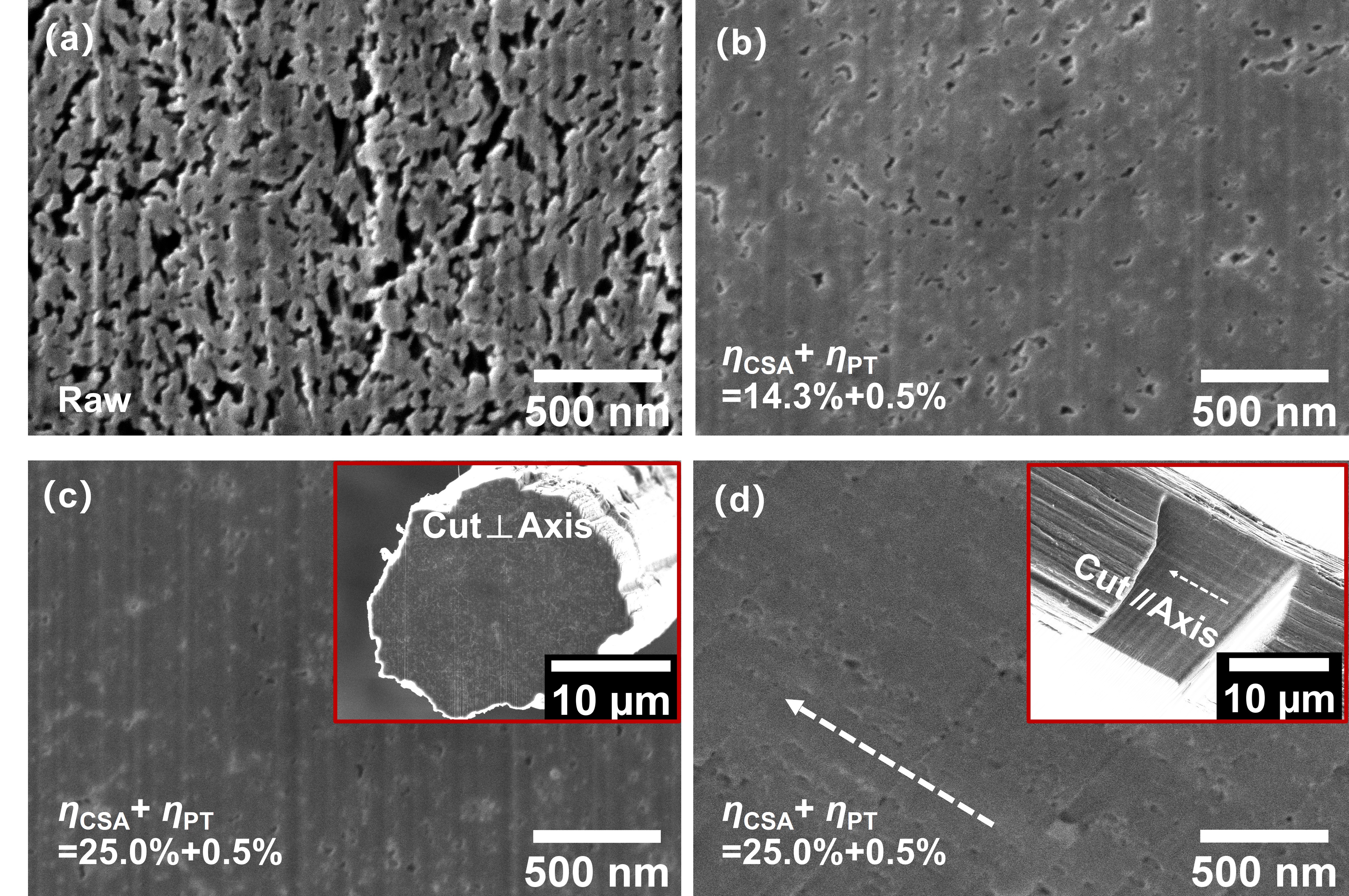}
    \caption{The evolution of porosity within CNTF after various levels of Double-Drawing. (a-c) On the cross-section cut perpendicular to the fiber axis, the porosity monotonically decreases with the increase of draw-ratio. (c) For the fully DD-CNTF, only fine voids can be found. (d) On the cross-section cut parallel to the fiber axis as shown in the inset, the remaining voids are all in a configuration of bead-chain along fiber axis (white arrow) instead of randomly distributed, indicating their presence between large bundles and aerogel layers.}\label{Fig:3}
\end{figure}

The evolution of voids within DD-CNTF was monitored by SEM of the cross-section cut by Focused Ion Beam (FIB). As shown in Fig.\ref{Fig:3}a-c, with increasing draw-ratio, the porosity decreases and reaches minimum on the fully DD-CNTF with fine voids (more details can be found in Fig.S2). Because the area surrounded by the voids is an indicator of the cross-section of a bundle, the ever-increasing solid area also illustrates the thickening of bundles. We further checked the cross-section cut parallel to the fiber axis (Fig.\ref{Fig:3}d). The remaining voids are in bead-chain configuration, indicating that the remaining voids in the fully DD-CNTF originate from the gaps between thick bundles and layers of aerogel.\par

\subsection{Enhancement of Mechanical Properties for DD-CNTFs}\label{Enhancement of Mechanical Properties for DD-CNTFs}
\begin{figure}[H]%
    \centering
    \includegraphics[width=0.9\textwidth]{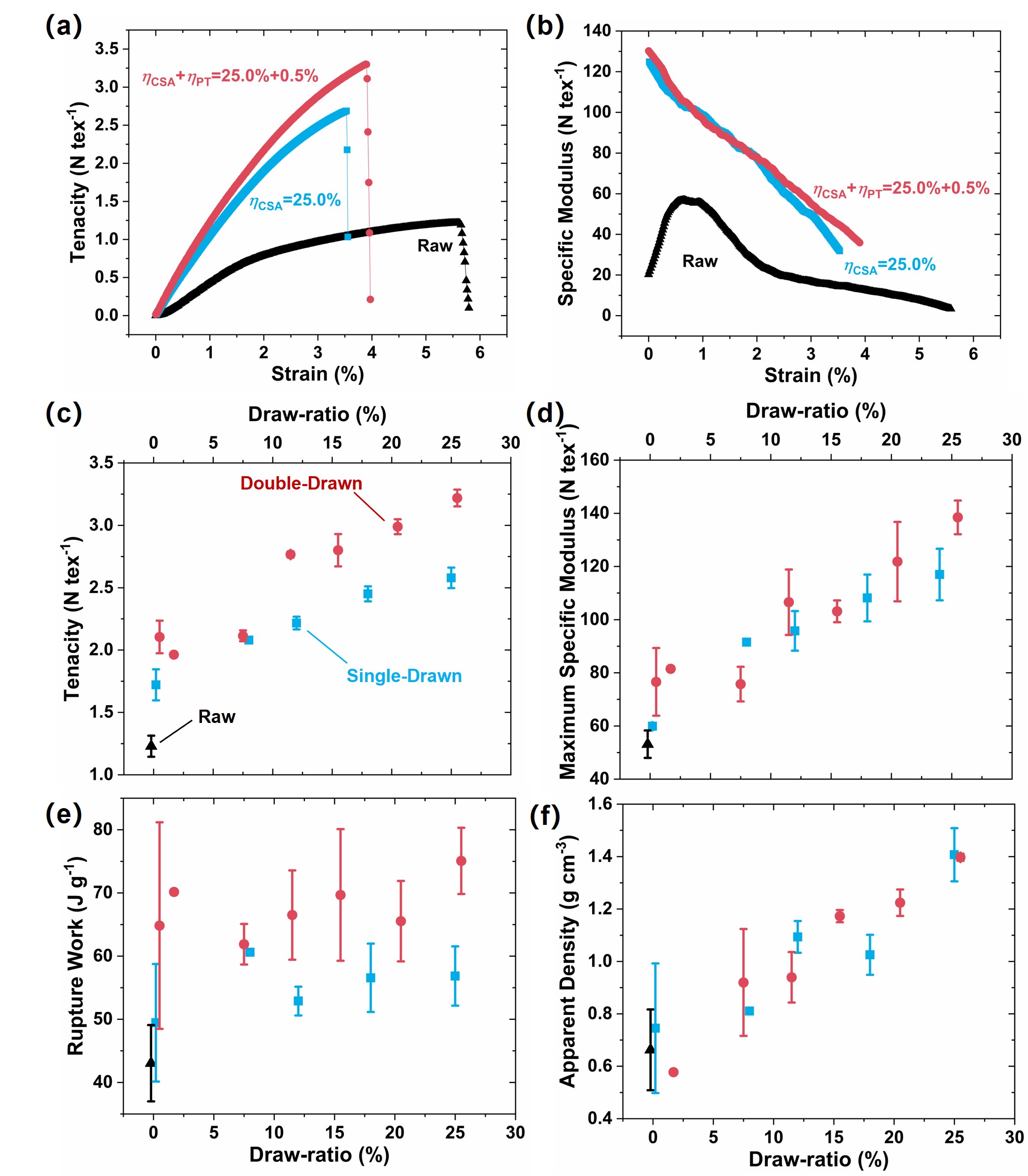}
    \caption{The evolution of the mechanical properties of CNTF with various draw-ratios. (a) In the representative tenacity-strain ($\varepsilon$) plots of CNTFs, for Double-Drawn (red dots, DD) and Single-Drawn (blue squares, SD) CNTFs, they are both significantly stiffened compared with the raw fiber (black triangles). With additional Poisson Tightening, the breaking tenacity further improved by 23\% without sacrificing ductility. (b) In the corresponding tangent specific modulus-strain plots, after drawing in CSA, the tangent specific modulus dropped monotonically with the increase of strain. (c) With CNTFs after different drawing, the evolution of the breaking tenacity and (d) the initial specific modulus versus the draw-ratios ($\eta{\rm_{CSA}}$ + $\eta{\rm_{PT}}$ are shown. Both DD and SD CNTF tenacity and the initial modulus monotonically increase with the increasing draw-ratio. (e) The work of rupture after Poisson Tightening increases, particularly for the CNTF above threshold drawing. (f) With the organized microstructure, the apparent density increases monotonically with the draw-ratio and reaches $\rm1.40\,g\,cm^{-3}$ for the fully DD-CNTF.}\label{Fig:4}
\end{figure}

The strength of fibers increase with the increasing density where the cross sectional-area diminishes for a given fiber as it is densified.\cite{Stallard2018} The cross-sectional area is poorly defined for porous nanomaterials and assemblies with ambiguous cross-sections or non-uniform diameters. Thus, to best characterize DD-CNTFs ``tenacity", i.e. ``specific strength" with units of $\rm N\,tex^{-1}$ (or $\rm GPa \ SG^{-1}$), is a well-defined indicator of load and does not have the ambiguity of absolute ``strength" with units of GPa. Tenacity is widely used for textile fibers and can be calculated directly by dividing stretching force with linear density ($LD$, mass/length), both of which can be unambiguously measured for fibers and porous nanomaterials.\par
As shown in Fig.\ref{Fig:4}a, compared with the raw CNTFs (black triangles), after full Double-Drawing (red dots), the fibers exhibit an increase in breaking tenacity from $\rm1.22\,N\,tex^{-1}$ to $\rm3.30\,N\,tex^{-1}$. When compared with the CNTF only drawn in CSA (Single-Drawn, SD-CNTF, blue squares in Fig.\ref{Fig:4}), the tenacity of the fully DD-CNTF has 23\% greater tenacity without any ductility degradation, which illustrates the importance of the Poisson Tightening process. Fiber strength, when accounting for SEM-measured cross sectional areas, increases from raw fiber strength of 0.9 GPa to 4.6 GPa for fully DD-CNTFs (see Fig.S3).\par
With the gradual increase in the draw-ratio, both DD-CNTFs and SD-CNTFs show a monotonic increase in tenacity (Fig.\ref{Fig:4}c). We find no saturation plateau on tenacity after threshold drawing as was shown with alignment (Fig.\ref{Fig:2}c). The further enhancing effect from Poisson Tightening is more obvious when $\eta{\rm_{CSA}}\textgreater10\%$. Furthermore, because of the consistency of ductility before and after the Poisson Tightening (Fig.S4), the work of rupture, i.e. energy absorbed during the rupture process increased to $\rm\sim70\,J\,g^{-1}$ for DD-CNTFs, from $\rm\sim42\,J\,g^{-1}$ for the raw CNTF, and $\rm\sim55\,J\,g^{-1}$ for SD-CNTFs.\par
In the corresponding tangent specific modulus-strain plots (M-S plots, Fig.\ref{Fig:4}b), for both DD-CNTFs and SD-CNTFs, the modulus reduces during the tensile testing process. The rising of modulus for the raw CNTF at the beginning comes from the deformation of the iCNT network.\cite{Park2019} For CNTFs after different draw-ratios, the initial modulus (maximum modulus) monotonically increases with the rising of draw-ratio (Fig.\ref{Fig:4}d) and reaches $\rm130.2\,N\,tex^{-1}$ for the specific modulus of fully DD-CNTF, compared with $\rm56.8\,N\,tex^{-1}$ for the raw CNTF. The close stacking and collapsed cross-section (Fig.S5) increases the CNTF apparent density from $\rm0.66\,g\,cm^{-3}$ of the raw fibers to $\rm1.40\,g\,cm^{-3}$ of the fully DD-CNTF (Fig.\ref{Fig:4}f), which is very close to the theoretical density of ideally compacted CNTs bundle ($\rm1.54\,g\,cm^{-3}$).\cite{Aliev2010,Behabtu2013}\par
It is also interesting that by only immersing in CSA, the CNTF gains $\rm \sim40\%$ increase in tenacity (Fig.\ref{Fig:4}c), along with short and straight CNT bundles on the fiber surface (Fig.\ref{Fig:1}d). We believe these phenomena originate from the spontaneously rearrangement of iCNTs, due to their high stiffness and persistence length (More discussion in S3).\cite{Fakhri2009,Yakobson2006}\par

\subsection{Enhancement of Thermal Properties for DD-CNTFs}\label{Enhancement of Thermal Properties for DD-CNTFs}
\begin{figure}[H]%
    \centering
    \includegraphics[width=0.9\textwidth]{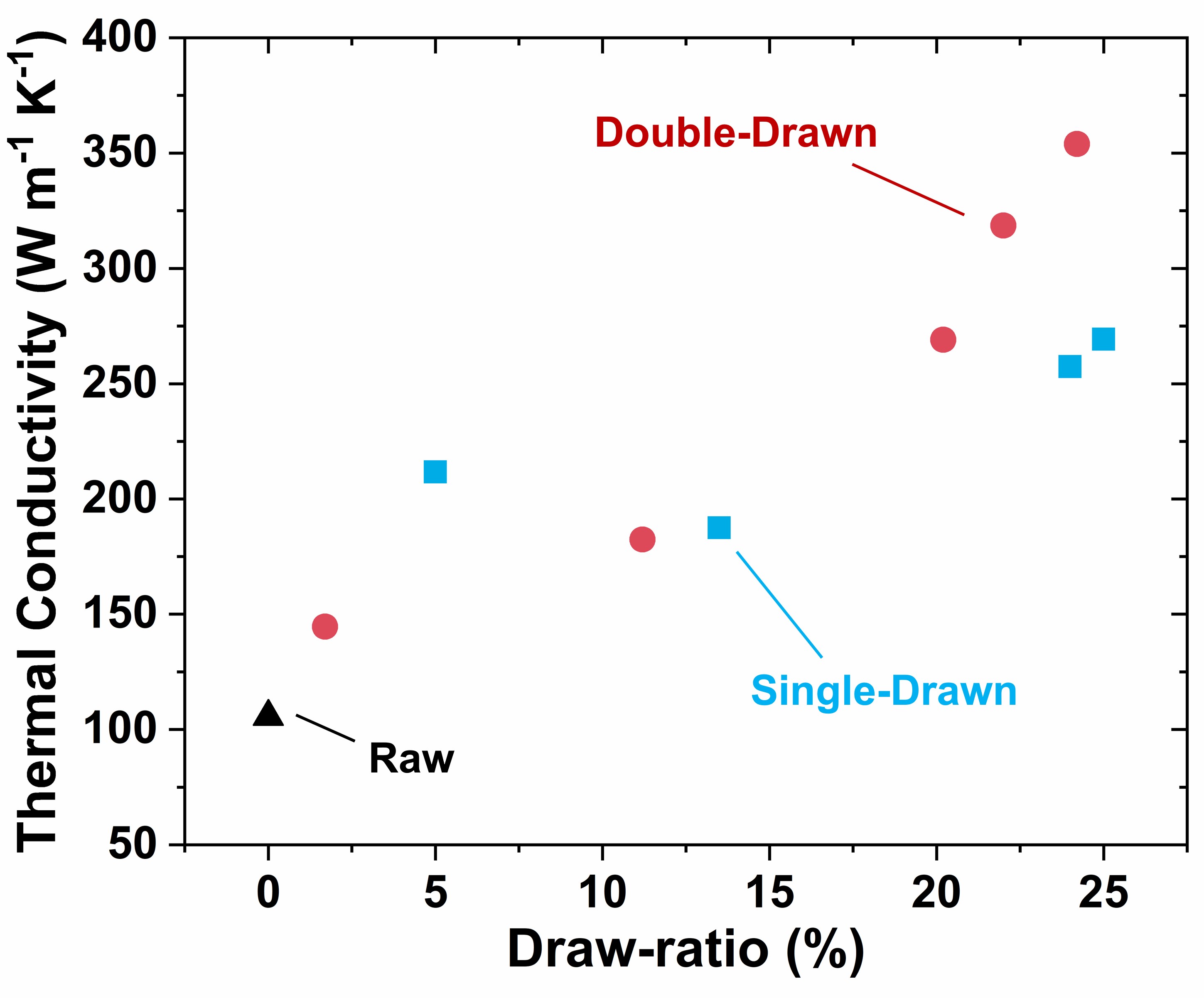}
    \caption{ Evolution of thermal conductivity with the increase of draw-ratio. Owing to the optimization of CNT-CNT junctions after drawing, the thermal conductivity of the fully DD-CNTF reaches $\rm354\,W\,m^{-1}\,K^{-1}$ which is 335\% higher than the raw CNTF, and 31\% higher than the fully SD-CNTFs.}\label{Fig:5}
\end{figure}

Since the CNT-CNT contacts (junctions) are the main source of thermal resistance\cite{Prasher2009} (e.g. phonon scattering centers\cite{Dresselhaus2000}), the fiber thermal conductivity also improves from the microstructure rearrangement after drawing. With the gradual increase of draw-ratio, the thermal conductivity increases monotonically, and the Poisson Tightening gives rise to further increase after the threshold drawing (Fig.\ref{Fig:5}). Consequently, thermal conductivity of the fully DD-CNTF reaches $\rm354\,W\,m^{-1}\,K^{-1}$ which is 335\% higher than the raw CNTF, and 31\% higher than the fully SD-CNTFs. Normalized by density, the specific thermal conductivity of the fully DD-CNTF ($\rm0.258\,W\,m^2\,K^{-1}\,kg^{-1}$) is five times higher than that of copper ($\rm0.044\,W\,m^2\,K^{-1}\,kg^{-1}$) and silver ($\rm0.041\,W\,m^2\,K^{-1}\,kg^{-1}$). The electrical conductivity also substantially increases from $\rm1650\,S\,cm^{-1}$ of the raw CNTF to $\rm10700\,S\,cm^{-1}$ of the fully DD-CNTF Fig.S6). While improved, the electrical conductivity remains below the conductivity of other bulk metal conductors, e.g., copper ($\rm\sim\!60000\,S\,cm^{-1}$).
Interestingly, in contrast to alignment evolution with the draw-ratio, the evolution of mechanical and conductive properties monotonically improves, which implies that there are other enhancing mechanisms besides the improvement in alignment.\par

\subsection{Comparison Between the DD-CNTF with Commercial Fibers}\label{Comparison Between the DD-CNTF with Commercial Fibers}
\begin{figure}[H]%
    \centering
    \includegraphics[width=0.9\textwidth]{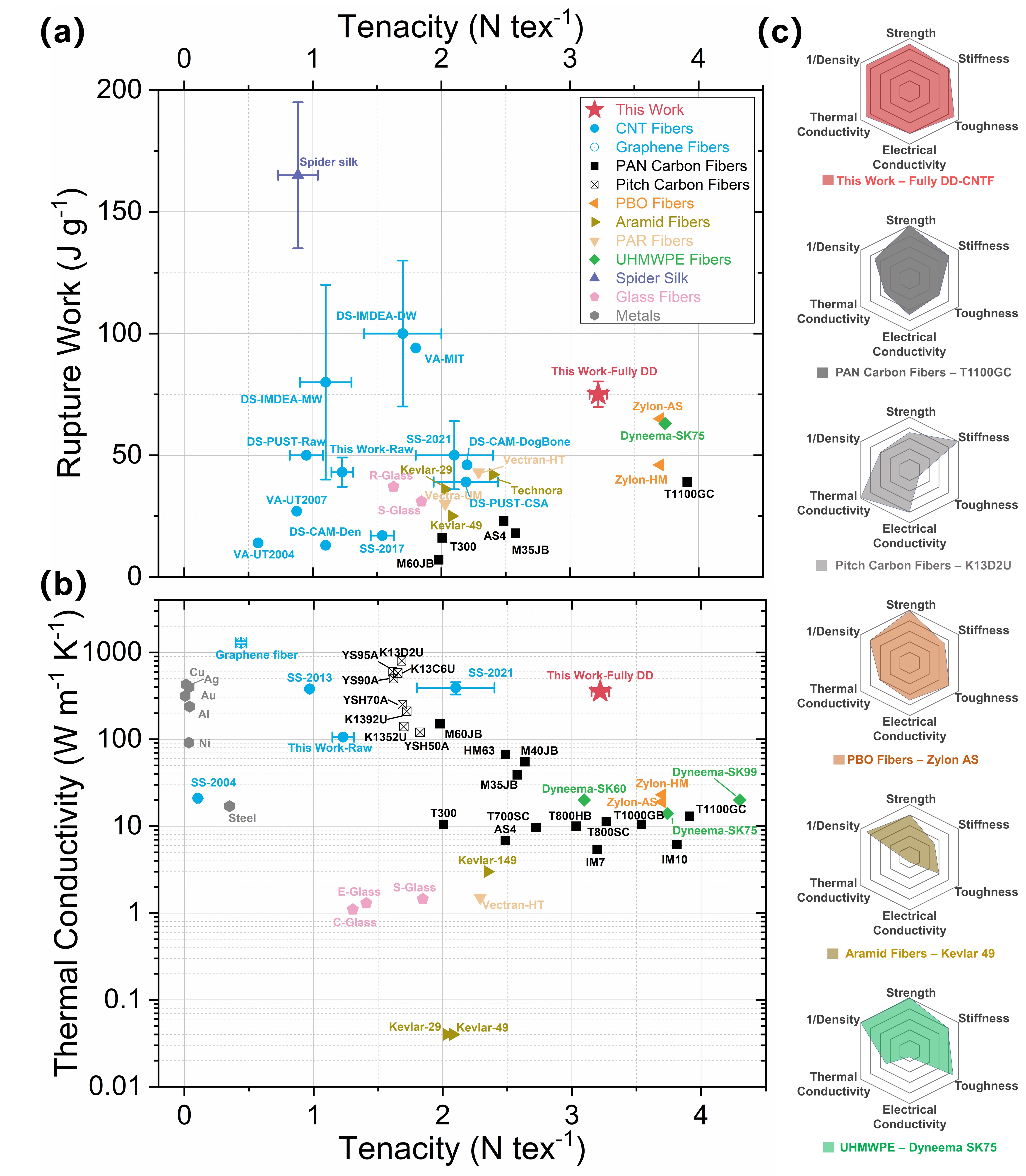}
    \caption{Properties comparison between CNTFs and commercial synthetic fibers. Ashby plots of (a) tenacity (specific tensile strength) versus work of rupture, and (b) tenacity versus thermal conductivity. The data points contain reported high-performance CNTFs (pure), high strength CFs (PAN-based CFs), highly thermal conductive CFs (Pitch-based CFs), some high-performance polymer fibers, and results from this work. (c) Radar plots comparing the performance of the fully DD-CNTF with PAN-based CF (T1100GC), Pitch-based CF (K13D2U), Zylon AS, Kevlar 49 and Dyneema SK75. The results are normalized by the maximum value of each characteristic (references in Table S1).}\label{Fig:6}
\end{figure}

As shown in Fig.\ref{Fig:6}, compared with current leading fibers, carbon nanotube fibers can exhibit overall versatile performance with a combination of high specific tensile strength (tenacity), work of rupture, thermal conductivity, and specific volume (e.g. low density). The fully DD-CNTFs have thermal conductivities that match the best pitch-based CFs and surpass them in terms of tenacity and ductility. The fully DD-CNTF tenacity is within 15\% of the strongest PAN-based CFs (T1000GB) and PBO fibers (Zylon AS and HM), while tougher and more conductive. The performance of fully DD-CNTFs are superior to Kevlar for all reported metrics highlighting potential applications of impact shielding or advanced structural usage with thermal management purposes.\par
The DD-CNTFs further the trend of the remarkable annual improvements of CNTF properties highlighted by Taylor et al over the last decade.\cite{Taylor2021} The DD-CNTFs is a method that enhances commercially-produced raw fibers with fiber test lengths (10-20 mm) from medium-grade crystalline materials ($I{\rm_G}:I{\rm_D}\sim5.3$, Fig.S7). When compared to congeneric fibers, including the solution-spun CNTF composed of higher crystallinity CNTs ($I{\rm_G}:I{\rm_D}=54 \ to \ 85$),\cite{Taylor2021} the fully DD-CNTF have improved strength and modulus, owing to the longer effective length of CNTs. Compared to other CNTF acid stretching processes, the DD-CNTF enables a larger $LD$ while improving the overall fiber properties. Short CNT strand (VA-MIT) have higher work of rupture than DD-CNTFs ($\rm94\,J\,g^{-1}$), but compromise strength and modulus with short fibers that were restricted lengths less than the individual CNTs ($\rm\sim1\,mm$).\cite{Hill2013} Critically, the present work enables sustained strength over gauge lengths that exceed the longest CNTs and offer a measure of scaled fiber performance that is comparable (0-25\% less) to high work of rupture of the IMDEA materials. Our materials do not require removal of catalyst or other impurities from CNTFs which account for $\rm7-8\,wt\%$ with marginally crystalline material.\par

\section{Discussions}\label{Discussions}
\subsection{Load Transfer Efficiency on Individual Tubes As Illustrated by the In-Situ Stretching Raman}\label{Load Transfer Efficiency on Individual Tubes As Illustrated by the In-Situ Stretching Raman}
\begin{figure}[H]%
    \centering
    \includegraphics[width=0.9\textwidth]{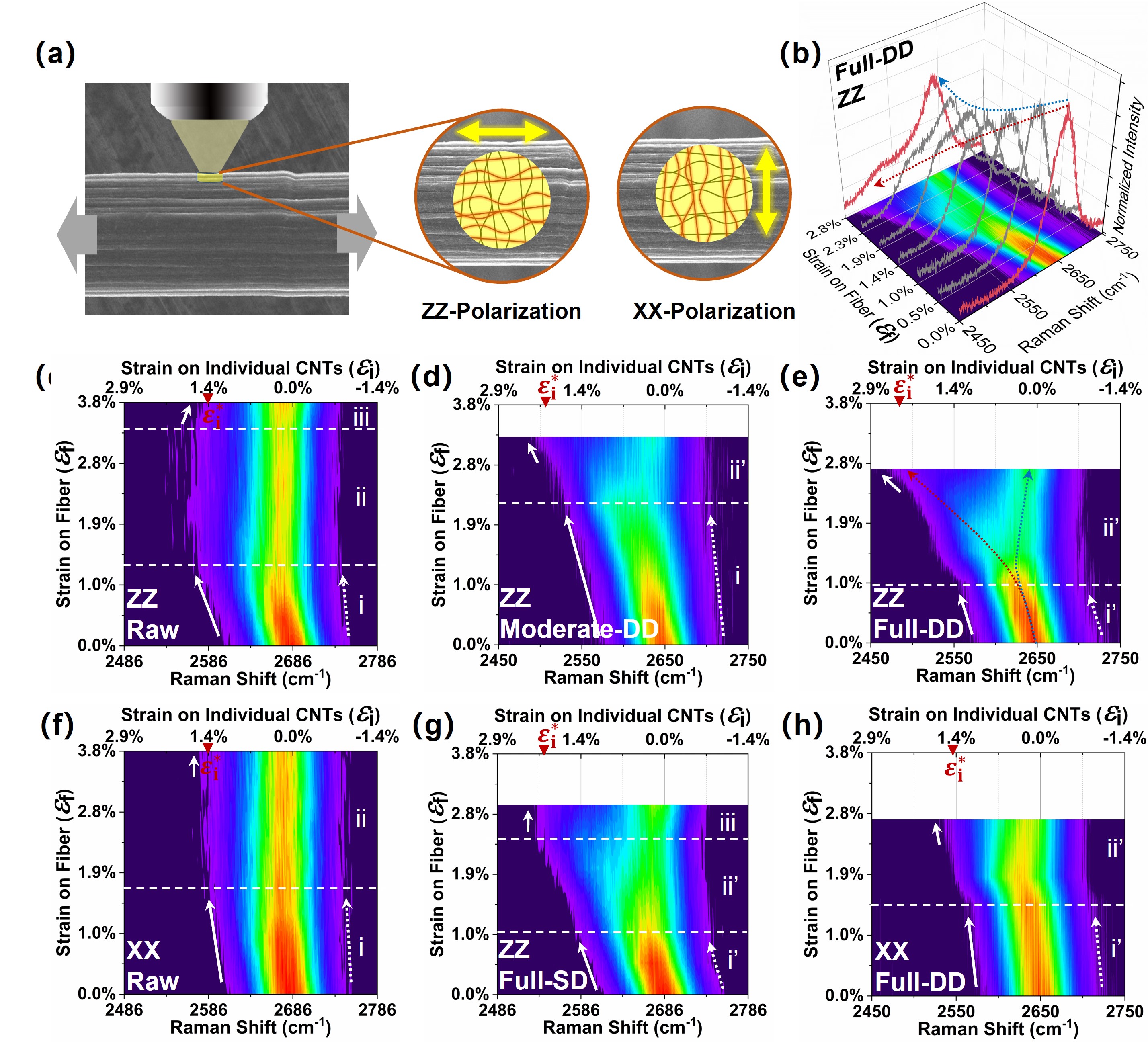}
    \caption{The Redshift of Raman G' mode of CNTFs evaluated by the In-Situ Stretching Raman (ISSR). (a) Schematic diagram of the ISSR characterization with polarized laser and ZZ / XX configuration. (b) For the fully DD-CNTF, with the increase of strain on CNTF ($\varepsilon\rm_f$), the redshift is distinctively efficient for portions of CNTs (indicated by red dotted arrow), which is indicative of increased of strain shared on iCNTs ($\varepsilon\rm_i$). Other portions of CNTs only partially redshift and return to the free state quickly (indicated by blue dotted arrow). The evolution of $\varepsilon\rm_i$ with $\varepsilon\rm_f$ are shown in the contour plots which respectively depict the evolution of the parallel iCNTs within the (c) raw CNTF, (d) the moderate DD-CNTF, (e) fully DD-CNTF, and (g) fully SD-CNTF, as well as the evolution of the perpendicular iCNTs within the (f) raw CNTF and (h) fully DD-CNTF. All the contour plots are normalized by the same scales, and divided into zones (Roman numerals) based on the different behaviors of redshift. The white arrows are guides to the redshifts, with their corresponding load transfer ratio ($\varepsilon{\rm_i}/\varepsilon{\rm_f}$) being elucidated by the arrows' slope. The maximum strains for iCNTs ($\varepsilon\rm_i^*$) are also marked.}\label{Fig:7}
\end{figure}

The distinctive increase in mechanical properties after the Double-Drawing process warrants a mechanistic study of the enhancing mechanisms to determine whether the DD-CNTFs have taken the full advantage of iCNTs' properties. Here we use the In-Situ Stretching Raman (ISSR) with the polarized detecting configuration to study the load transfer efficiency on iCNTs. In contrast to the common polarized Raman characterization that solely depicts the CNTs alignment in fibers, ISSR enables assessment of the distribution of strain on iCNTs.\cite{Chang2010} In ISSR, the C-C bonds softens under stretching, redshifting the Raman G' mode proportionally with strain.\cite{Cronin2004,Mohiuddin2009}\par
It is important to note that the ISSR spectrum is an amalgamation of every section on thousands of iCNTs within the laser spot, which can experience various strain. Along each CNT the strain can be distributed unevenly. Based on the antenna effect of CNTs,\cite{Saito2011,Zhang2015b} we use ZZ/XX polarization configuration to detect the strain distribution of iCNTs parallel/perpendicular to the axis of CNTF (Fig.\ref{Fig:7}a). As shown in Fig.\ref{Fig:7}b, when CNTF strain ($\varepsilon\rm_f$) is gradually increased, sections in the spectrum start to redshift differently (red and blue dotted lines), from which the corresponding strain distribution among iCNTs ($\varepsilon\rm_i$) can thus be deduced (more details can be found in S4). To highlight the evolution of $\varepsilon\rm_i$ with corresponding $\varepsilon\rm_f$, we accumulate these spectra of each CNTF into its contour plot (Fig.\ref{Fig:7}c-h).
As the fiber strain increases to $\varepsilon\rm_f$ =1.27\% (Zone i) the tubes parallel to the raw CNTF axis (ZZ configuration, Fig.\ref{Fig:7}c) exhibit a small broadening in the spectrum tail of low-frequency (white solid arrow), while the middle and high frequency sections change little (white dotted arrow). This implies that only a small proportion of CNTs share the strain on the fiber, while the others do not participate.\cite{Cronin2005,Kumar2007}) When $\varepsilon\rm_f$ =1.27\% the average $\varepsilon\rm_i$ can be deduced $\rm \sim0.22\%$ (calculation details can be found in S4). We use load transfer ratio $LTR\equiv\varepsilon{\rm_i}/\varepsilon{\rm_f}$ as a figure of merit for load sharing. Thus, the average $LTR$ ($\overline{LTR}$) is only $\rm \sim0.17$, which indicates that the fiber strain $\varepsilon\rm_f$ is primarily a result of straightening of curled tubes, alignment of tubes towards axis direction or the relative slippage among tubes, rather than the strain increase on iCNTs. For larger strain, $\varepsilon\rm_f\textgreater1.27\%$, the redshift reaches a plateau without any changes in the spectrum (Zone ii). $\varepsilon\rm_i$ does not further increase and reaches a maximum, $\varepsilon\rm_i^*\sim1.65\%$ (the corresponding $LTR^*$ is $\rm \sim0.49$), indicating the occurrence of slippage between iCNTs. Approaching the failure point ($\varepsilon\rm_f\geq3.38\%$, Zone iii), the broadening disappears (short white dotted arrows), which indicates the remaining small portion of CNTs break and return to the initial state without any strain.\par
As a comparison, the fully DD-CNTF (Fig.\ref{Fig:7}e) shows a significant enhancement on the load sharing on the iCNTs. As load increases to $\varepsilon\rm_f$ =0.97\%, the entire peak of G' mode redshifts without obvious broadening (white dotted arrow vs white solid arrow), which we referred as Zone i'. The corresponding $\overline{LTR}$ increases to $\rm \sim0.35$ at $\varepsilon\rm_f$ =0.97\%. The redshift of the entire G' mode indicates that a major portion of CNTs take the load with fiber. More importantly, with greater strain, $\varepsilon\rm_f\textgreater0.97\%$, there is no plateau of redshift in Zone ii'. Instead, the spectrum splits into 2 groups: one group (indicated by the red arrow) continues to share the load as $\varepsilon\rm_f$ increases, while the other group (indicated by the blue arrow) gradually releases when strained, indicating the successive failure of thick bundles. For the former group, $\varepsilon\rm_f$ continues to increase and finally achieves $\varepsilon\rm_i^*\sim2.44\%$ just before the rupture of CNTF, The corresponding $LTR^*$ can then be deduced as high as 0.89, indicating that a portion of CNTs synchronize with the fiber to take the load from the beginning until the fracture. We did not observe the Zone iii strain releases to the initial state as occurred with the raw CNTF.\par
For the moderate DD-CNTF ($\eta{\rm_{CSA}}=10.4\%$ and $\eta{\rm_{PT}}=0.5\%$, Fig.\ref{Fig:7}d), its behavior falls in between raw CNTF and fully DD-CNTF. The $\overline{LTR}$ at the end of Zone i is $\rm \sim0.23$. Like the fully DD-CNTF, Zone ii' appears with $\varepsilon\rm_f$ increases. $\varepsilon\rm_i^*$ finally reaches $\rm \sim2.07\%$ just before the rupture of CNTF. The corresponding $LTR^*$ is $\rm \sim0.64$. Without the Poisson Tightening process, the fully SD-CNTF behaves similarly in Zone i' and ii' (Fig.\ref{Fig:7}g), with $\overline{LTR}\sim0.29$ at the end of Zone i', and $LTR^*$ is $\rm \sim0.83$ when $\varepsilon\rm_i^*\sim2.09\%$. Importantly, in the Zone iii where $\varepsilon\rm_f$ approaches the rupture point, $\varepsilon\rm_i^*$ cannot further increase, while the other portion of CNTs release to the initial state free of strain. As a contrast, for the tubes perpendicular to the fiber axis (XX-Polarization), for both raw (Fig.\ref{Fig:7}f) and the fully DD-CNTFs (Fig.\ref{Fig:7}h), during the whole period, $\varepsilon\rm_i$ changes very little with the increase of $\varepsilon\rm_f$, indicating the redundance of tubes perpendicular to the axis.\par
Collectively the ISSR results for CNTFs with different enhancing process, enables insights into CNTFs and the Double-Drawing process, namely (1) only a small portion of CNTs in the raw CNTF can take axial load; (2) with only small strain on iCNTs, slippage will happen in raw CNTF; (3) by drawing in CSA, a larger portion of CNTs can participate in load sharing upon initial loading; (4) by drawing in CSA, higher strain on iCNTs are needed to lead to a slippage; (5) Poisson Tightening can further increase the strain capacity on iCNTs. The cumulative impact of these optimization from the Double-Drawing elucidates the substantial increase in the breaking tenacity of the fully DD-CNTF.\par

\subsection{Enhancing Mechanism - the Increase of Effective Bundles and the Extension of CNTs Effective Length in Bundle}\label{Enhancing Mechanism - the Increase of Effective Bundles and the Extension of CNTs Effective Length in Bundle}
\begin{figure}[H]%
    \centering
    \includegraphics[width=0.9\textwidth]{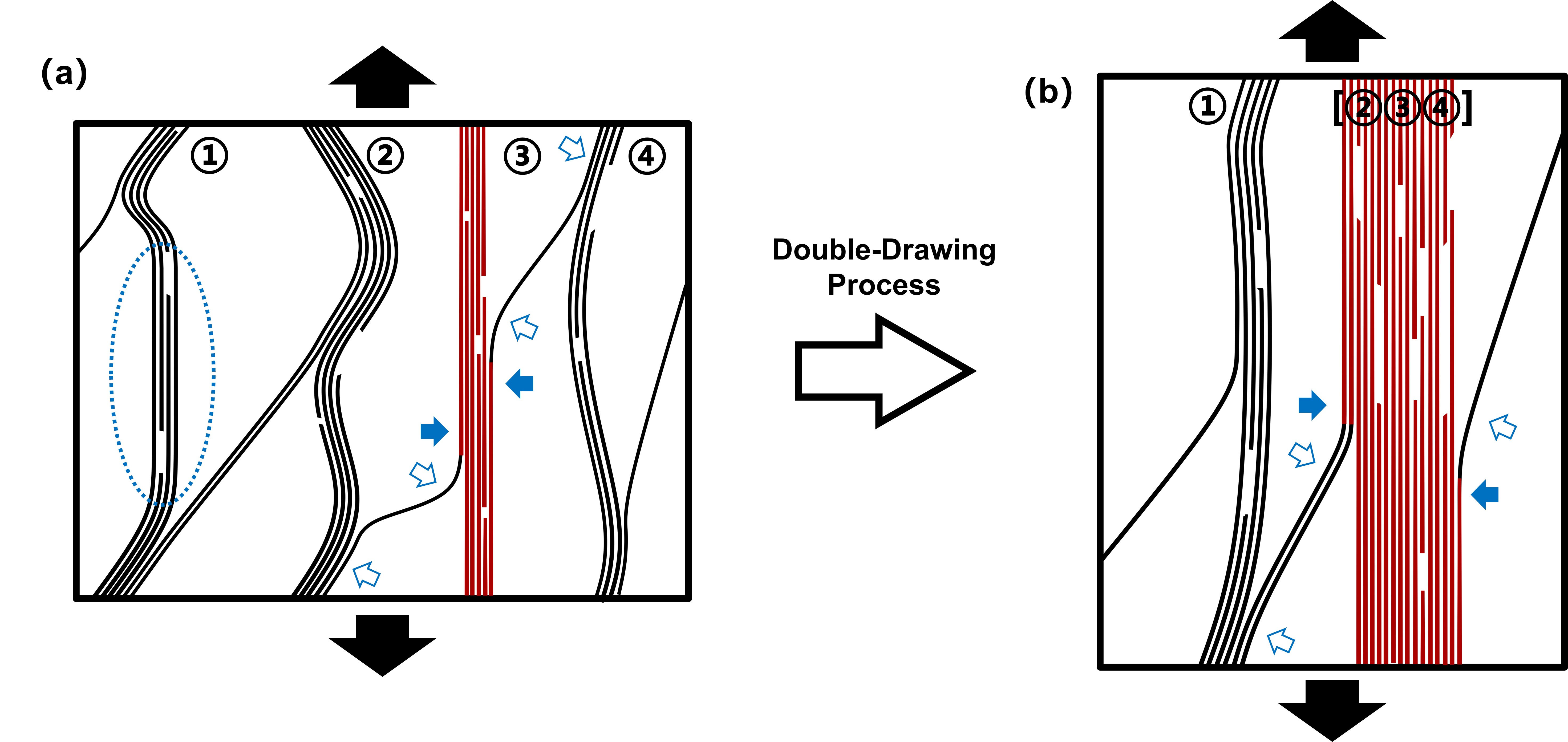}
    \caption{The 2D schematic graph of the optimization on bundles being effective and the effective length of tubes within a bundle after Double-Drawing process. (a) in the disordered CNTs network, tubes (lines) aggregate as bundles (marked as \textcircled{1} to \textcircled{4}). Within the schematic cell under vertical stretching, only shortest bundle connecting the vertical surfaces takes load, i.e., being effective (painted in red). Tubes outside the loaded bundle remain idle despite portions being oriented parallel to the load (circled with blue dotted circle). For tubes that link to multiple bundles, only the length section of tubes attached to an effective bundle take load (solid blue arrows), while the external length section remains idle (hollow blue arrows). (b) With the Double-Drawing process, the crumpled iCNTs are straightened as the shortest pathway and more tubes are compressed into a large bundle ([\textcircled{2}\textcircled{3}\textcircled{4}]), extending the effective length section.}\label{Fig:8}
\end{figure}

The improvement of properties after the saturation of alignment optimization, and the ever rising of $\overline{LTR}$ and $\varepsilon\rm_i^*$ after processing, indicates additional factors must be included to develop a representative mechanism for CNTF loading. Misorientation of monomer units along the polymer chain has been recognized as the primary factor that leads to the reduction of stiffness of many synthetic fibers.\cite{Adams1987} The tensile Young's modulus of fibers is commonly estimated by the appropriate average of the moduli of all monomers along the axis.\cite{Hearle2008} However, the assumption implies that all monomers evenly participate within a fiber, which is not suitable for a fibrillar assembly like the CNTF. The movement and deformation of iCNTs within CNTFs are not all affine. Instead under load a tensioning line frequently appears from the disordered network in raw CNTFs, which indicates the stress concentration, also the only portion of CNTs bearing the load (detailed analysis can be found in S5).\par
As shown within a simplified CNTF cell (Fig.\ref{Fig:8}a), if the load is exerted on the vertical surfaces of the cell (along axis), there is no medium between CNTs to transmit the load, thus only the shortest CNT bundle is loaded (bundle \textcircled{3}, the red lines). While alignment analysis by WAXD shows a substantial alignment, the portion of idle CNTs cannot be determined. Therefore, for the fibrillar structures the straightening of bundles is a more indicative metric for fiber strength than orientation factor. After the drawing in CSA, more crumpled tubes and bundles straighten ([\textcircled{2}\textcircled{3}\textcircled{4}]) and become effective to link the ``shortest" distance (the red lines in Fig.\ref{Fig:8}b). They jointly participate in sharing the load after the initial stretching. More correlations between this factor to the optimization of ISSR results and mechanical properties can be found in S5.\par
Nevertheless, only accounting for the increased fraction of load-bearing bundles in network does not fully explain the much higher $\varepsilon\rm_i^*$ for the DD-CNTF. As the $\varepsilon\rm_i^*$ appears immediately prior to failure, the much-improved tenacity for the fully DD-CNTF must therefore be considered. For the elastic interface (static friction), the stress exerted on iCNTs $\sigma{\rm_i}=Y\,\varepsilon\rm_i$, is balanced by the friction from the surrounding tubes, where $Y$ and $\varepsilon\rm_i$ is the tube's Young's modulus and strain, respectively.\cite{Gupta2020} When $\sigma\rm_i$ increases to the critical value $\sigma\rm_i^*$, slippage will occur and the elastic interface will begin to deform plasticly. Immediately prior to slippage, $\sigma{\rm_i^*}\,A{\rm_t}=f{\rm_s}\,L{\rm_{eff}}$, where $A\rm_t$ is the cross-sectional area of the tube, $f\rm_s$ is tube's maximum static friction coefficient per unit length, and $L{\rm_{eff}}$ is the effective length of tube that shares the load (friction). Thus, $\sigma\rm_i^*$ is the maximum value for $\sigma\rm_i$ and the corresponding $\varepsilon\rm_i^*$ can be deduced by:\par

\begin{equation}\label{eq1}
\varepsilon{\rm_i^*}=\frac{\sigma{\rm_i^*}}{Y}=\frac{f{\rm_s}\,L{\rm_{eff}}}{Y\,A{\rm_t}}
\end{equation}

In a fibrillar structure, particularly raw CNTF (entire length of iCNTs $L\rm\!\sim100\,\mu m$\cite{Mikhalchan2019}), tubes within the hierarchical network only partly align with any specific bundle, and may be incorporated into many bundles. Because the load can only be transmitted through the coupling between adjacent tubes, only the tube section attached to an effective bundle can participate in sharing the load (the red lines as indicated by solid blue arrows in Fig.\ref{Fig:8}), i.e., $L{\rm_{eff}}\gg L$ (Fig.\ref{Fig:1}b-c). With the Double-Drawing process, longer length of tubes aggregate into the effective bundles ([\textcircled{2}\textcircled{3}\textcircled{4}] in Fig.\ref{Fig:8}b), with $L{\rm_{eff}}$ approaching $L$. Consequently, $\varepsilon\rm_i^*$ needed to activate the slippage also increases, which delays the failure of DD-CNTF and improves the tenacity (discussed further in S5).\par
Although the conduction pathway for heat is not essential to be shortest, the thermal conduction benefits when more bundles and larger fraction of tubes become effective. The junction boundaries between bundles are likely a first order contribution to thermal resistance,\cite{Chalopin2009} which become sparser along the fiber axis after the Double-Drawing. With the increase of $L{\rm_{eff}}$ and alignment of adjoining bundles, the joint length and interfacial area between tubes are extended, improving the thermal conductance through the corresponding junctions.\par

\section{Conclusions}\label{Conclusions}
In summary, we have developed a technique to enhance CNT fibers by successively drawing the raw direct-spun CNTFs within CSA and chloroform at ambient temperature and pressure, which aligns the CNT bundles, densifies the fibers and removes the residue CSA via Poisson Tightening and rinsing. With full Double-Drawing, CNTFs reach a tenacity of $\rm3.30\,N\,tex^{-1}$ (4.60 GPa), Young's modulus of $\rm130\,N\,tex^{-1}$, rupture work of $\rm70\,J\,g^{-1}$, electrical conductivity of $\rm10700\,S\,cm^{-1}$, and thermal conductivity of $\rm354\,W\,m^{-1}\,K^{-1}$. The resulting simultaneous optimization properties results in an attractive overall performance, continuing the impressive line of improvement seen within CNTFs worldwide in recent years. We anticipate further advancement in material properties using the Double-Drawing technique which is applicable to other CNTs ensemble including forests, films, fibers, and aerogels.\par
Importantly, in addition to the known dependence of properties on CNT alignment and stacking, new evidence of the load transfer coefficient on individual CNTs highlights the importance of (i) straightening of CNT bundles which increases the proportion of effective bundles jointly sharing the load and (ii) the higher barrier of slippage activation within bundles, which originates from the effective tube length increase within effective bundles.\par
The comprehensive improvement of CNTF properties enables long-sought applications of advanced load-bearing fibers that can simultaneously dissipate heat, shield impacts, or dissipate electrical charge. Comprehensive understanding, advancement of these multitude of mechanisms, and resulting properties are necessary to manifest bulk CNT fiber properties that approach the horizon of properties offered by individual CNTs.

\section{Experimental/Methods Section}\label{Experimental/Methods Section}
\textit{CNT fibers preparation}: Continuous CNTFs were fabricated using the floating catalyst method,\cite{Li2004,Motta2007a} and supplied by Tortech Nanofibers Ltd. The produced CNT aerogels from a tube furnace were mechanically pulled out, densified by acetone, and spun continuously winded. Although a small tension force is applied during the spinning process to obtain a preferential alignment along fiber axis, the anisotropic ratio is always within 0.85.\cite{Zhang2018}\par
\textit{Enhancing CNTF with the Double-Drawing process}: (i) \textit{Immersing and first drawing}: The raw CNTF is fixed at its lower end inside a dropping funnel, and its upper end fixed on a spin rotor. The CNTF is straightened out but without a pre-tension. After being immersed in chlorosulfonic acid (CSA), CNTF is firstly drawn to a specific ratio ($\eta{\rm_{CSA}}$; (ii) \textit{Poisson Tightening}: After the immersing solvent being changed into chloroform, the CNTF is immediately further drawn by $\eta{\rm_{PT}}$. (iii) \textit{Rinsing}: after the Double-Drawing processes, the CNTF is successively rinsed in water and acetone, and finally vacuum dried.\par
\textit{Linear Density measurement}: The $LD$ is measured based on Direct Single-fiber Weight Determination, following ASTM D1577-07(2018) OPTION B. The weight of a CNTF with length $\rm\sim100\,mm$ is measured with a Sartorius SE2 Ultra-micro balance. We find it worthy to notice the importance of accurately measurement of $LD$, because the susceptibility of fiber's tenacity to $LD$. The frequently used Vibroscopic method in reports is abandoned here, because of the potential serious underestimation of $LD$, if the ``stiffness correction" was overlooked (ASTM D1577-07(2018) OPTION C - the standard for the Vibroscopic method). More discussion can be found in S2.\par
\textit{Tensile test}: The CNTFs are tested with the Single-Fiber Testers (Textechno FAVIMAT with Load cell of 210 cN and delicately aligned clamps ($\rm4\,mm$ hard rubber), the force resolution of $\rm\sim0.0001\,cN$, the displacement resolution of $\rm0.1\,\mu m$). The CNTFs are tested with gauge length of $\rm10\,mm$, stretching speed of 1 mm min-1 and pretension of $\rm0.1\,cN\,tex^{-1}$. Every sample is tested for 3 specimens to guarantee the repeatability of results. Stretching speeds of 0.2, 2 and 5 $\rm mm\,min^{-1}$, and gauge length of $\rm20\,mm$ have been tried to generate similar results.\par
\textit{Thermal and Electrical conductivity measurement}: The thermal conductivity of CNTF along fiber axis is performed with a homemade measuring apparatus based on a self-heating method.\cite{Zhou2017} The electrical conductivity of CNTFs along fiber axis is measured in air at room temperature (1 atm, $\rm25-27\,^{\circ}$C, relative humidity: $\rm40\pm3\,\% RH$) by a homemade testing stage using the four-electrode method and steady-state method.\cite{Zhou2016}\par
\textit{In-Situ Stretching Raman}: The suspended CNTFs are ends fixed onto a manual stretching stage to detect the Raman signal with HORIBA HR800 micro-Raman spectroscopy. We excite the Raman G' mode with linearly polarized laser, and only collect the scattered radiation in the parallel polarization with a Glan Polarizer, so that only iCNTs with their axis close to parallel with the laser polarization can be detected. For the ZZ/XX configuration,\cite{Damnjanovic1999} the polarizations of both incident and scattered photons are parallel/perpendicular to the axis of CNTFs, offering the strain distribution of CNTs along/normal to the fiber axis;\par
\textit{Wide Angle X-ray Diffraction (WAXD)/Small-Angle X-ray Scattering (SAXS)}: The CNTFs with different processing are studied using a small and wide-angle diffractometer (Molecular Metrology SAXS system) equipped with a sealed microfocus tube (MicroMax-002+S) emitting Cu $\rm K\alpha$ radiation (wavelength of $\rm 0.1542\,nm$), two G\"{o}bel mirrors, and three pinhole slits. CNTFs with diameter of $\rm 18-50\,\mu m$ were suspended onto a holder perpendicular to the beam and measured at ambient temperature. All the raw data are analyzed by SAXSGUI. For data analysis of WAXD, the sharp equatorial reflections at $\vec{q}\sim1.8\,{\rm\AA ^{-1}}$ corresponding to the scattering from (002) reflection of the inter-layer spacing of a few walled CNT, and to a higher order reflection of the hexagonal packing of parallel CNTs. Both possibilities are due to the planes perpendicular to the CNT axis. To obtain the azimuthal profile of (002) scattering, the intensity is integrated in the range of $\rm1.6-1.9\,\AA ^{-1}$. With the increase of the alignment, two peaks emerge in the azimuthal profile around the preferred alignment, from which the Full Width Half Maximum is obtained. For data analysis of SAXS, the integrating range is $\rm0.04-0.1\,\AA ^{-1}$ to obtain the azimuthal profile of scattering.\par
\textit{Other characterization}: The cross-sections of CNTFs are fabricated with FIB (FEI Helios 600i). The cross-section is firstly cut with Gallium ions with current of 9 nA (30 kV) and then finely polished under current of 0.79 nA, after which the SEM of cross-sections are conducted by electron beam of FIB. The SEM for other CNTFs is conducted on a TESCAN MIRA3. HRTEM is conducted on an FEI Talos F200X TEM working under 80 kV to reduce the damage to CNTs.\par

\bmhead{Supplementary materials}\label{Supplementary materials}\par
Supplementary materials are available.\par

\bmhead{Acknowledgements}\label{Acknowledgements}
\textbf{Funding}: This work is supported by EPSRC project 'Advanced Nanotube Application and Manufacturing (ANAM) Initiative' [grant numbers: EP/M015211/1]. WY Zhou and HP Liu thank the partial support by the National Key R\&D Program of China [grant numbers: 2018YFA0208402, 2020YFA0714700]. 
\textbf{General}: The authors especially thank Ms. Mingzhao Wang, Prof. Wei Tan, Dr. Joe Stallard, Dr. Le Cai, Ms Rulan Qiao, Dr. Sarah Stevenson, for their kind support and useful discussion. The author also thanks Dr. Heather Greer, Dr. Rafail Khalfin, Prof. Yachin Cohen, and Prof. Thurid S. Gspann for TEM, WAXD, SAXS, and mechanical experiment support. 
\textbf{Competing interests}: All authors declare that they have no competing interests. 
\textbf{Data and materials availability}: All data needed to evaluate the conclusions in the paper are present in the paper and/or the Supplementary Materials. Additional data related to this paper may be requested from the authors.
\textbf{Author Contributions}:
Conceptualization: XZ, MDV, SSX, AMB; 
Methodology: XZ, MDV, WBZ; 
Formal analysis: XZ, WBZ, AMB; 
Investigation: XZ, WBZ, JTP, AK; 
Resources: WBZ, LI, AK, JTP, FRS, ZBW, JAE; 
Validation: MDV, WBZ, WYZ, JAE, AMB; 
Supervision: MDV, HPL, WYZ, SSX, AMB; 
Project administration: YCW, AMB; 
Funding acquisition: HPL, WYZ, AMB; 
Writing—original draft: XZ, MDV, AMB; 
Writing—review \& editing: XZ, MDV, WBZ, XJW, JTP, HPL, WYZ, JAE, AMB.\par

\bibliography{library}

\end{document}

\title[Supplementary material]{\textbf{Supplementary material} \par \setlength{\parskip}{20pt} Simultaneously Enhanced Tenacity, Rupture Work, and Thermal Conductivity of Carbon Nanotubes Fibers by Increasing the Effective Tube Contribution}


\author[1,2]{\fnm{Xiao} \sur{Zhang}}\email{zhangx@iphy.ac.cn}

\author*[2]{\fnm{Michael} \sur{De Volder}}\email{mfld2@cam.ac.uk}

\author[3]{\fnm{Wenbin} \sur{Zhou}}\email{wbzhou@bjut.edu.cn}

\author[2]{\fnm{Liron} \sur{Issman}}\email{li242@cam.ac.uk}

\author[1]{\fnm{Xiaojun} \sur{Wei}}\email{weixiaojun@iphy.ac.cn}

\author[4]{\fnm{Adarsh} \sur{Kaniyoor}}\email{ak2011@cantab.ac.uk}

\author[4]{\fnm{Jerónimo Portas} \sur{Terrones}}\email{jt451@cam.ac.uk}

\author[2]{\fnm{Fiona} \sur{Smail}}\email{frs25@cam.ac.uk}

\author[1]{\fnm{Zibo} \sur{Wang}}\email{wangzibo@iphy.ac.cn}

\author[1]{\fnm{Yanchun} \sur{Wang}}\email{ycwang@iphy.ac.cn}

\author[1]{\fnm{Huaping} \sur{Liu}}\email{liuhuaping@iphy.ac.cn}

\author[1]{\fnm{Weiya} \sur{Zhou}}\email{wyzhou@iphy.ac.cn}

\author*[4]{\fnm{James} \sur{Elliott}}\email{jae1001@cam.ac.uk}

\author*[1]{\fnm{Sishen} \sur{Xie}}\email{ssxie@iphy.ac.cn}

\author*[2]{\fnm{Adam} \sur{Boies}}\email{amb233@cam.ac.uk}

\affil[1]{\orgdiv{Beijing National Laboratory for Condensed Matter Physics}, \orgname{Institute of Physics, Chinese Academy of Sciences}, \city{Beijing}, \postcode{100190}, \country{China}}

\affil[2]{\orgdiv{Department of Engineering}, \orgname{University of Cambridge}, \city{Cambridge}, \postcode{CB2 1PZ}, \country{UK}}

\affil[3]{\orgdiv{MOE Key Laboratory of Enhanced Heat Transfer and Energy Conservation, Beijing Key Laboratory of Heat Transfer and Energy Conversion}, \orgname{Beijing University of Technology}, \city{Beijing}, \postcode{100124}, \country{China}}

\affil[4]{\orgdiv{Department of Materials Science and Metallurgy}, \orgname{University of Cambridge}, \city{Cambridge}, \postcode{CB3 0FS}, \country{UK}}

\maketitle
\section*{S1. Experiment details on the Double-Drawing process.}\label{S1}

The raw CNTFs are continuously fabricated with a CVD reactor at $\rm1100-1200\,^{\circ}$C using the floating catalyst method,\cite{Li2004,Motta2007a} and supplied by Tortech Nanofibers Ltd. In the fabrication process, a preheated feedstock consisting of toluene (carbon source), ferrocene (catalyst precursor), thiophene (promoter), and hydrogen (carrier gas) with flow rates of $\rm 1-2\,g\,h^{-1}$, $\rm 100-400\,mL\,min^{-1}$, $\rm 10-20\,mL\,min^{-1}$, and $\rm 1000-2000\,mL\,min^{-1}$, respectively, was injected into a reactor to form CNT aerogels continuously. The CNT aerogels were mechanically pulled out, densified by acetone, and spun continuously with a motorized winding system at a winding rate of $\rm 15\,m\,min^{-1}$. Although a small tension force is applied during the spinning process to obtain a preferential alignment along fiber axis, the anisotropic ratio is always within 0.85.\cite{Zhang2018} The raw CNTFs, normally with linear density $\rm \sim 0.5\,tex$ and $I{\rm_G}:I{\rm_D}\sim 5.3$ are used for the subsequent processing.
\par
To enhance the raw CNTFs, the fiber is firstly fixed at its lower end inside a dropping funnel, and its upper end fixed on a spin rotor. After being immersed in chlorosulfonic acid (CSA) for 30 s, CNTF is drawn to a specific ratio ($\eta{\rm_{CSA}}$) at a rate of $\rm 1\,mm\,s^{-1}$. To fully draw the CNTF, the maximum ratio $\eta{\rm_{max}}$ is used, which is slightly less than the failure ratio $\eta\rm^*$ in CSA. The drawing rates from $\rm 0.1\,mm\,s^{-1}$ to $\rm 10\,mm\,s^{-1}$ have been tried without generating obvious difference. 
\par
After the immersing solvent being changed into chloroform, the drawn CNTF is immediately further drawn by $\eta{\rm_{PT}}=0.5\%$. When the CNTF is only immersed in chloroform, pimples of CSA always appear on the surface of drawn CNTF, this is because the closely compacted tubes on the outer fiber layer hinders the diffusion of the remaining CSA from the inner layers. The remaining CSA keeps screen the vdW forces through which the load transfer. Therefore, the second drawing process in chloroform is necessary. The pimples keep disappearing during the second drawing process.
\par
After the Double-Drawing processes, the CNTF is successively rinsed in water and acetone, and finally vacuum dried at $\rm 200\,^{\circ}$C for 2 hours. 
\par
We find the strong dependence of $\eta\rm^*$ on linear density ($LD$) of CNTF. For example, for CNTFs with $LD \rm \sim 0.47 \,tex$, $\eta\rm^* \sim 28\%$; and for CNTFs with $LD \sim 5.38\,tex$, $\eta\rm^* \sim 125\%$. With the thickening of the CNTF, larger $\eta\rm^*$ is needed to straighten all the tubes across the thickness. Thus, $\eta\rm^*$ can be estimated as $\eta\rm^*=41 \ \times \ LD^{1/2} [\%\,tex^{-1/2}]$. The empirical relationship should be also applicable for other direct spun CNT assemblies, like mat and film. 
\par
In our experiment, all the raw CNTFs are used directly without any purification. Although by heating and acid treatment, impurities like catalysts and amorphous carbon can be removed, we find that the inevitable defects generated during the purifying process result in the excessive drawing at variable points and the degradation of the draw-ratio on other part.
\par

\section*{S2. Precise Determination of Linear Density for CNTFs.}\label{S2}
The accuracy of tenacity (specific strength) value for micro-fibers heavily relies on the accurate measurements of linear density ($LD$). 
\par
Researchers usually used the Vibroscopic methods to deduce $LD$ of CNTFs. By introduce transverse vibration with oscillatory force, $LD$ can be deduced by finding the fundamental resonance frequency ($f_0$) of the fiber under known conditions of gauge length ($l$) and pre-tension tension ($F\rm _{PT}$) with following equation:\cite{Montgomery1952}
\par

\begin{equation}\label{eq:S1}
    LD=\frac{F\rm _{PT}}{4 f_0^2 l^2}
    \tag{S1}
\end{equation}

\par
However, Equation (\ref{eq:S1}) only works well under the assumption that tested fibers are perfectly flexible. However, the assumption can cause problems for CNTFs. As mentioned in ASTM D1577-07(2018) OPTION C, correction must be made on $f_0$ to account for the “stiffness effect”,\cite{Hearle2008,Montgomery1952,Voong1953} following the equation below:
\par

\begin{equation}\label{eq:S2}
    LD_c=\frac{F\rm _{PT}}{4 f_0^2 l^2}(1+2\alpha+5.47\alpha^2) = \frac{F\rm _{PT}}{4 f_0^2 l^2}(1+\delta )
    \tag{S2}
\end{equation}

\par
Here, $f_{0c}$ is the measured apparent fundamental resonant frequency, stiffness factor $\alpha \equiv \sqrt{4 E I / l^2 F_{PT}}$, and $E$ is the Young's modulus of the fiber, $I$ is the inertia moment of the fiber about the neutral axis. For a circular cross-section fiber with diameter, $I = \pi d^4/64$. And $\delta \equiv 2\alpha + 5.47\alpha ^2$ is the increase factor for $LD$ because of the “stiffness effect”. Thus, the real tenacity consequently should be $1/(1+\delta)$ lower compared to that without considering “stiffness effect”. 
\par
Therefore, $\delta$ becomes non-negligible if fibers possess (a) high Young's modulus, or (b) relative short gauge length, or (c) being tested with low pre-tension. As to the post-processed CNTFs with high modulus and commonly used testing parameters, the measured $LD$ always deviates away from that based on perfect flexible string model. Based on some typical reported data, $\delta$ deduced can be as high as $33\%$, which is unreasonable to be ignore. 
\par
Considering the above factors, we choose the Direct Single-fiber Weighing method as the reliable measurement of $LD$ following ASTM D1577-07(2018) OPTION B, because the accuracy is only dependent on the accuracy with which the fibers can be length measured and weighed.
\par

\section*{S3. Surface morphology and tenacity of CNTFs only immersed in CSA.}\label{S3}
As shown in Fig.4c, by only immersing in CSA, the CNTF gains $~40\%$ increase in tenacity. Moreover, after CSA processing, CNTFs always contain short and straight CNT bundles on the fiber surface (Fig.4d). We believe these straight bundles originate from the high stiffness and persistence length of iCNTs and their bundles which behave similarly to semiflexible filaments, with their persistence length much larger than the pristine curved diameter formed in raw fiber.\cite{Fakhri2009,Yakobson2006} When the athermal structure of CNTF “melts” upon immersion in CSA, the curved iCNTs spontaneously rearrange towards an equilibrium organized structure with improved properties, while dangling iCNTs on the surface rearrange into a rod-like shape as observed.
\par

\begin{figure}[H]%
    \centering
    \includegraphics[width=0.9\textwidth]{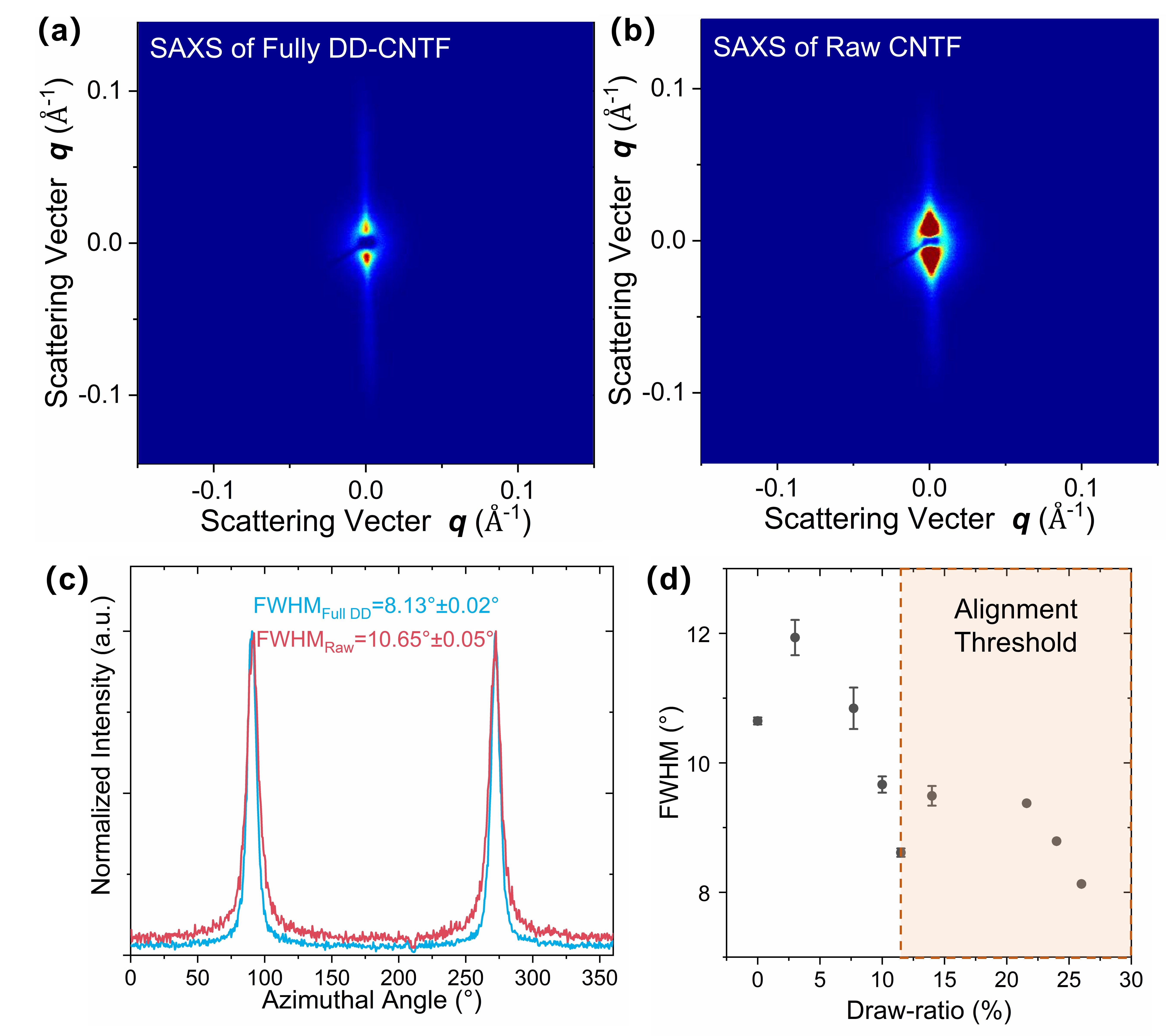}
    \caption{\textbf{SAXS pattern evolution of CNTFs with the increase of draw-ratio.} (a-b) the SAXS patterns from the suspended fully DD-CNTF and raw CNTF. (c) The azimuthal profiles of the SAXS intensity illustrate peaks from the preferred alignment in CNTF. (d) The evolution of alignment peak in SAXS azimuthal results of suspended CNTFs. Like the (002) peak of WAXD azimuthal profile, the alignment of iCNTs within CNTF indicated by the FWHM of peaks in SAXS azimuthal profile decreases with the gradually increase of draw-ratio but seems to saturate after threshold level of drawing ($12\%$)}\label{Fig:S1}
\end{figure}

\begin{figure}[H]%
    \centering
    \includegraphics[width=0.9\textwidth]{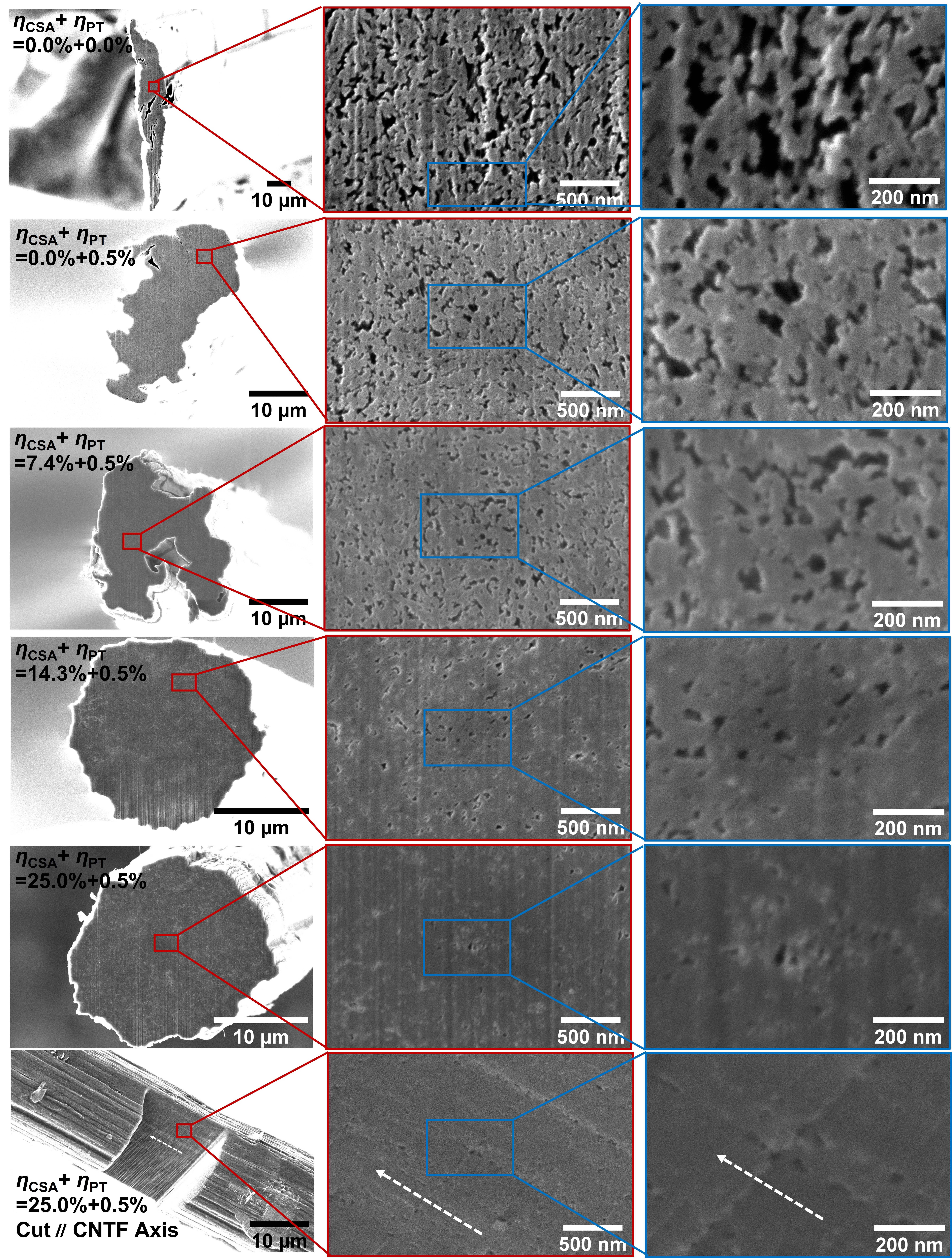}
    \caption{\textbf{The evolution of the porosity within the CNTFs after the increase of drawing.} On the cross-section cut with FIB, the trend is obvious that the porosity monotonically decreases with the increase of draw-ratio. We further checked the cross-section parallelly cut along the fiber axis (last row), the stubborn voids are all in a configuration of beads-chain along fiber axis (white arrow). Because the area surrounded by the voids is an indicator of the cross-section of a bundle, the ever increasing of solid area also illustrates the thickening of bundles, as mentioned in the Microstructure section of main text. The bright lines in images come from the Curtain Effect originated during the FIB cutting. We remove the bright lines in the last row by image processing method\cite{Loeber2017}to highlight the voids configuration.}\label{Fig:S2}
\end{figure}

\begin{figure}[H]%
    \centering
    \includegraphics[width=0.9\textwidth]{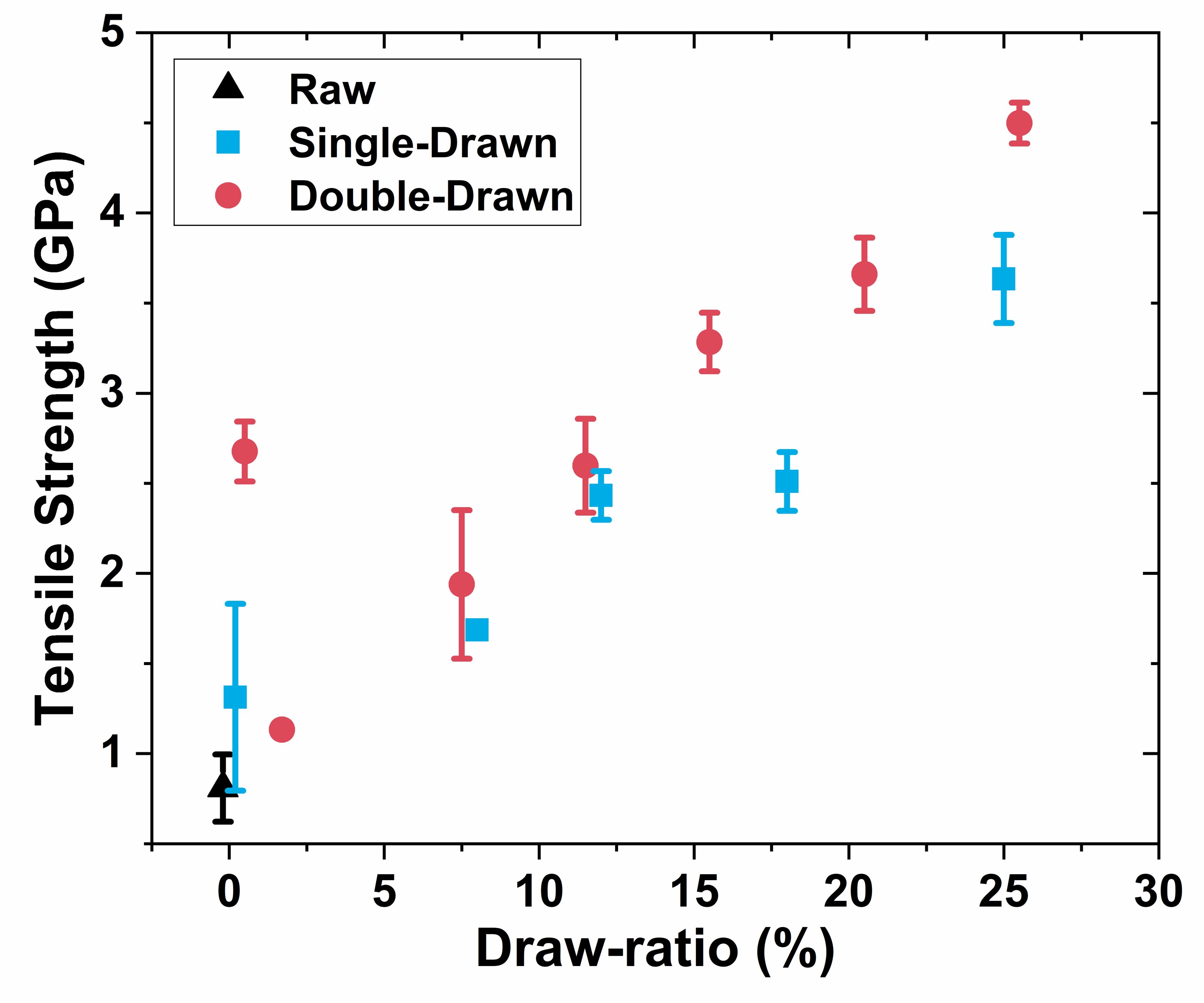}
    \caption{\textbf{The evaluation of the CNTFs' ultimate tensile stress (strength) with the rising of draw-ratio.} All the strength values are calculated by dividing the breaking force with the apparent cross-sectional area measured with SEM. For the fully DD-CNTF, the strength is improved to $\rm 4.50\pm 0.11\,GPa$, compared with $\rm 0.81\pm 0.19\,GPa$ for the raw CNTF.}\label{Fig:S3}
\end{figure}

\begin{figure}[H]%
    \centering
    \includegraphics[width=0.9\textwidth]{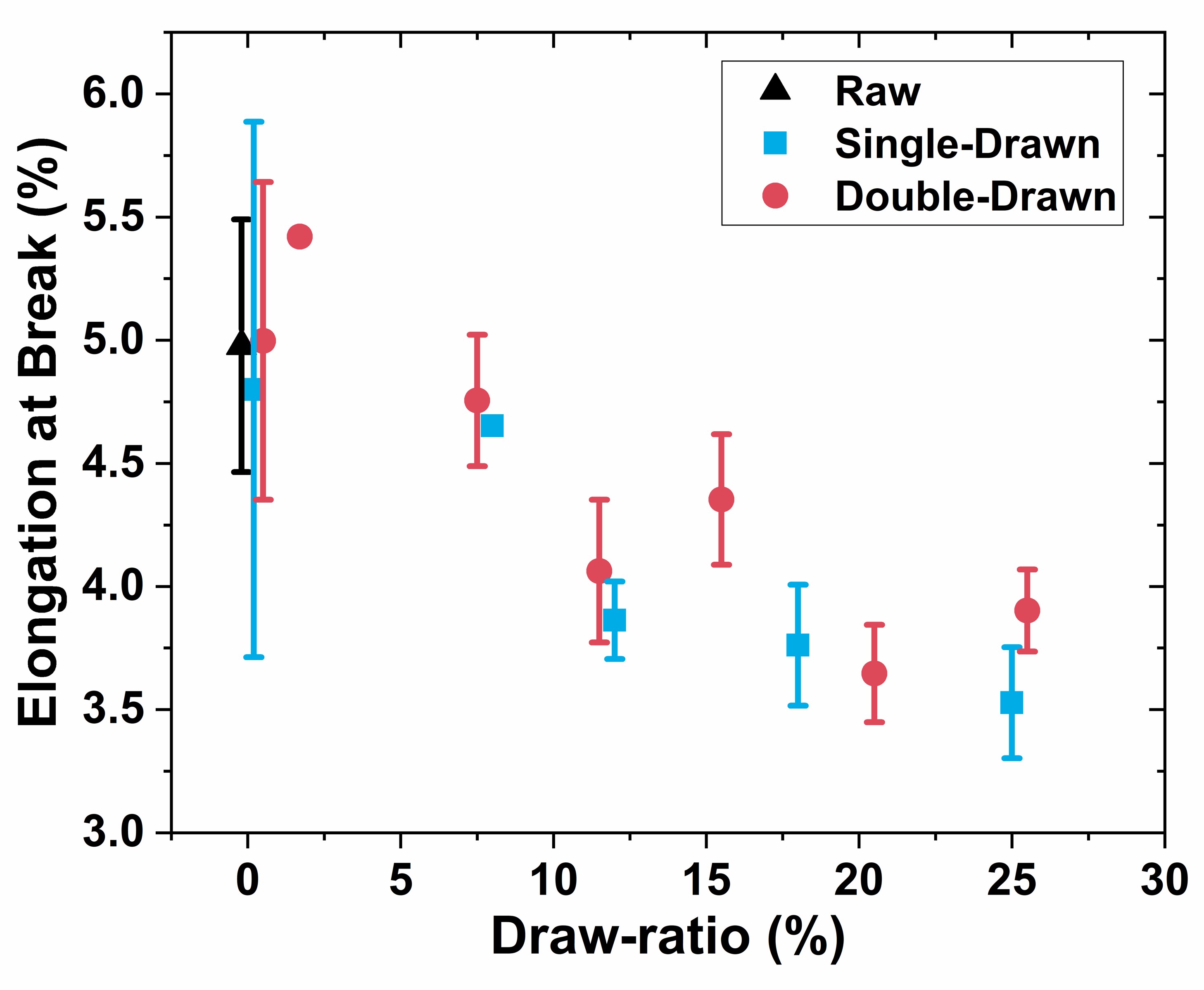}
    \caption{\textbf{The evaluation of the CNTFs' elongation at break (ductility) with the increase of draw-ratio.} For both Double-Drawn (DD) and Single-Drawn (SD) CNTFs, the ductility keeps decreasing with the draw-ratio. Although exerted extra drawing (Poisson Tightening), DD-CNTFs are not sacrificed on ductility when compared with SD counterparts. The large error bars for the raw and slightly drawn CNTFs come from the difference between fiber samples. The error bars are reduced with drawing.}\label{Fig:S4}
\end{figure}

\begin{figure}[H]%
    \centering
    \includegraphics[width=0.9\textwidth]{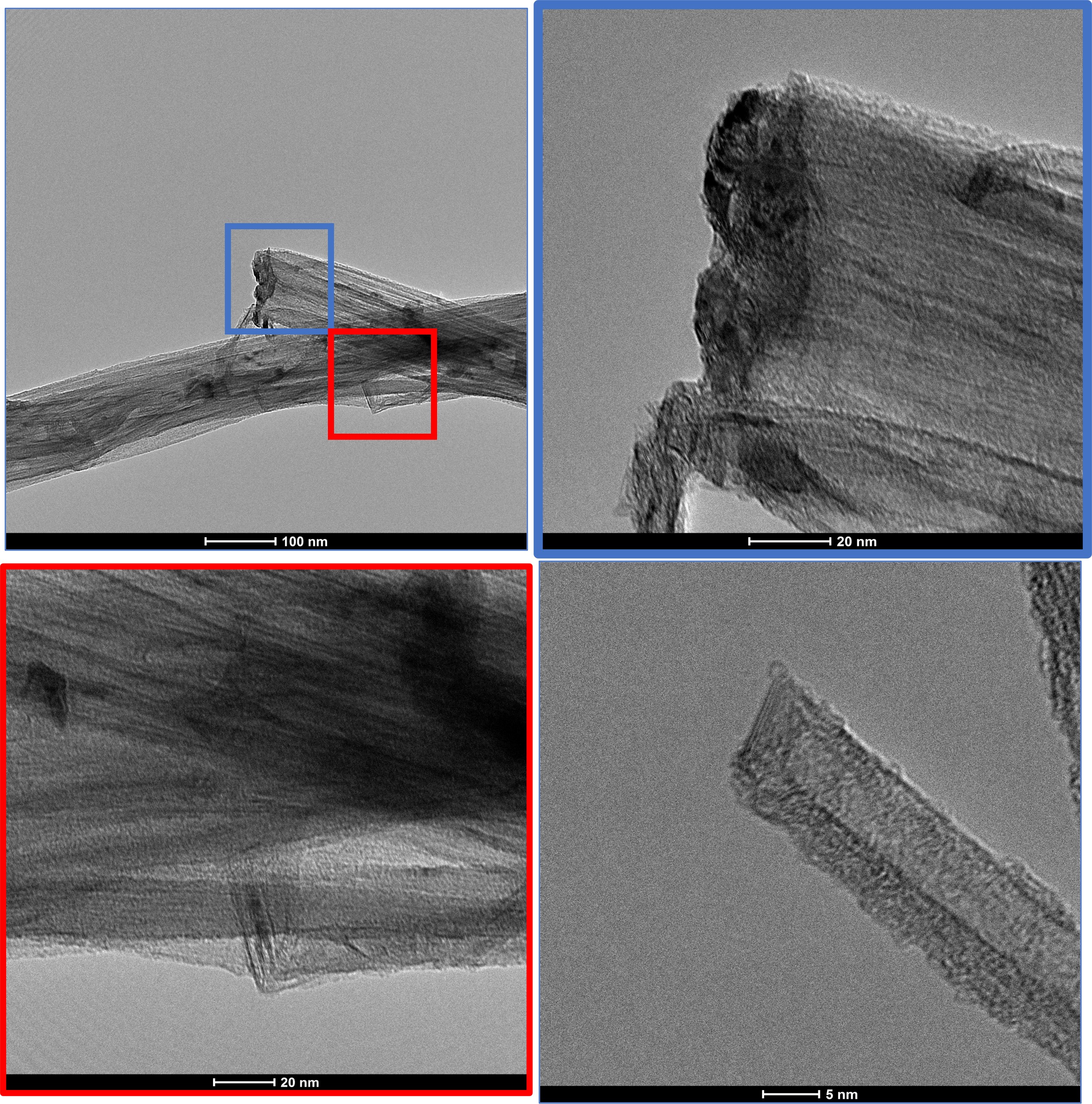}
    \caption{\textbf{With HRTEM characterization, all the cross-sections of bundles we found are collapsed ones instead of the deformed irregular ones.}\cite{Cho2018} We think it may come from the radial compressing during the Double-Drawing process. The collapsed cross-section also has a positive impact on $f \rm_s$ by maximizing the contact area between tubes, which can further enhance the bundle.\cite{Motta2007a}}\label{Fig:S5}
\end{figure}

\begin{figure}[H]%
    \centering
    \includegraphics[width=0.9\textwidth]{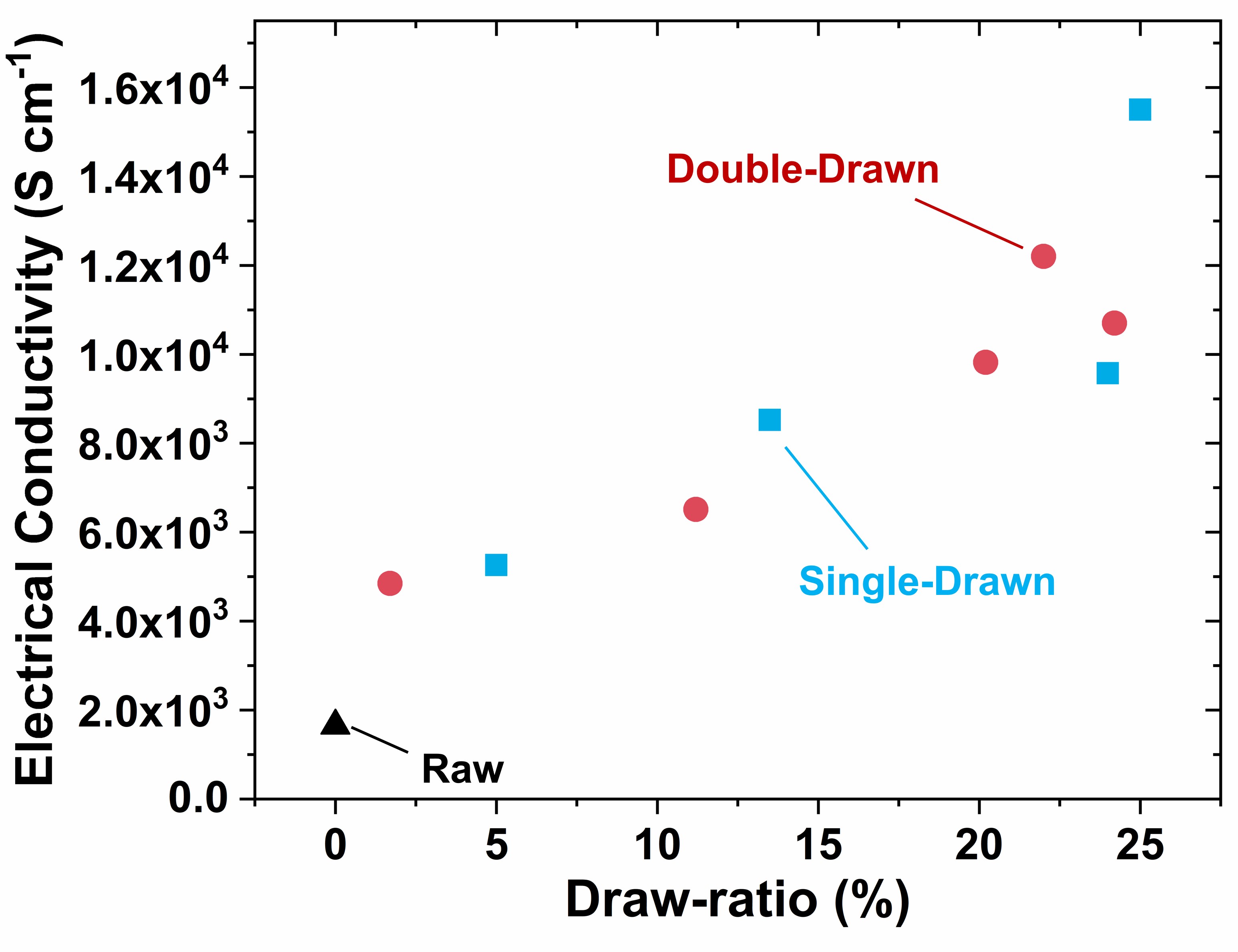}
    \caption{\textbf{Evolution of electrical conductivity with the increase of draw-ratio.} The electrical conductivity substantially increases from $\rm 1650\,S\,cm^{-1}$ of the raw CNTF to $\rm 10700\,S\,cm^{-1}$ of the fully DD-CNTF, and $\rm 15500\,S\,cm^{-1}$ of the fully SD-CNTF. However, unlike the thermal conductivity, the optimization of electrical conductivity may be attributed to doping effect, ordering on microstructure, and electron-electron interactions,\cite{Behabtu2013,Bulmer2021} further study is underway to elucidate the origin of the increase.}\label{Fig:S6}
\end{figure}

\begin{figure}[H]%
    \centering
    \includegraphics[width=0.9\textwidth]{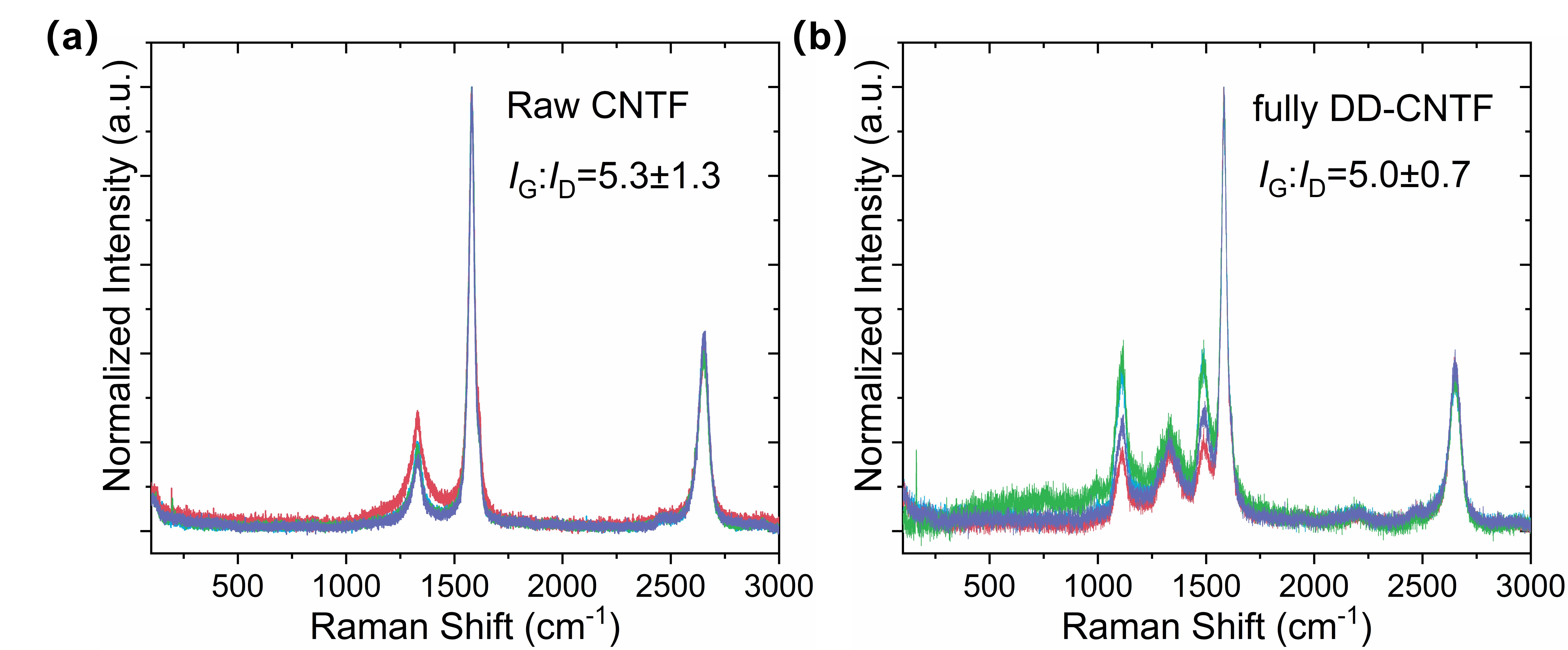}
    \caption{\textbf{Raman spectra of (a) raw CNTF and (b) fully DD-CNTF.} The high performance fully DD-CNTF Although the raw CNTFs are only medium-grade crystalline materials ($I{\rm_G}:I{\rm_D}=5.3\pm1.3$), they can still be processed to produce high performance DD-CNTFs. Moreover, as the $I{\rm_G}:I{\rm_D}$ for fully DD-CNTFs is $5.0\pm0.7$, we have not observed obvious increase of defects after the Double-Drawing process.}\label{Fig:S7}
\end{figure}

\begin{sidewaystable}
    \sidewaystablefn%
    \begin{center}
    \begin{minipage}{\textheight}
    \caption{Summary of mechanical, thermal, and electrical properties of pure CNT fibers in our and others' work, and some commercial high-performance fibers. The data list is also used for the Ashby plot of main text.}\label{Table:S1}
    \tiny
    \begin{tabular*}{\textheight}{@{\extracolsep{\fill}} c|c|c|c|c|c|c|c|c@{\extracolsep{\fill}}}
    \toprule%
    Name\footnotemark[1] & Commnets	
    &  \makecell{Density \\{} [$\rm g\,cm^{-1}$]}
    &  \makecell{Tenacity\\{}[$\rm N\,tex^{-1}$]}
    &  \makecell{Initial Modulus\\{}[$\rm N\,tex^{-1}$]}
    &  \makecell{Elongation\\{}[$\%$]}
    &  \makecell{Rupture Work\\{}[$\rm J\,g^{-1}$]}
    &  \makecell{Electrical\\{}Conductivity\\{}[$\rm MS\,m^{-1}$]}
    &  \makecell{Thermal\\{}Conductivity\\{}[$W\,m^{-1}\,K^{-1}$]}\\
    \midrule
    \rowcolor{cyan!50}This Work-Fully DD & Fully Double-Drawn CNTF & $1.40\pm0.01$ & $3.22\pm0.07$ & $138\pm6$ & $3.9\pm0.2$ & $75\pm5$ & $1.07$ & $354$ \\ \rowcolor{cyan!50}
    This Work-Raw & Raw Direct-spun CNTF & $0.66\pm0.15$ & $1.23\pm0.08$ & $53\pm5$ & $5.0\pm0.5$ & $43\pm6$ & $0.16$ & $106$ \\ \rowcolor{cyan!50}
    SS-2017\cite{Tsentalovich2017} & Solution-spun CNTF & $1.50\pm0.10$ & $1.54\pm0.09$ & $160\pm40$ & $1.8\pm0.2$ & $17$ & $8.5$ & $-$ \\ \rowcolor{cyan!50}
    SS-2004\cite{Ericson2004} & Solution-spun CNTF & $1.11\pm0.07$ & $0.10\pm0.01$ & $108\pm9$ & $-$ & $-$ & $-$ & $21$ \\ \rowcolor{cyan!50}
    SS-2013\cite{Behabtu2013,Tsentalovich2017} & Solution-spun CNTF & $1.30\pm0.10$ & $0.97$ & $92\pm38$ & $1.4\pm0.5$ & $-$ & $2.9$ & $380\pm15$ \\ \rowcolor{cyan!50}
    SS-2021\cite{Taylor2021} & Solution-spun CNTF & $1.93\pm0.16$ & $2.10\pm0.30$ & $-$ & $3.5\pm0.6$ & $50\pm14$ & $10.90\pm1.20$ & $391\pm63$ \\ \rowcolor{cyan!50}
    DS-IMDEA-DW\cite{Fernandez-Toribio2018} & Direct-spun collapsed DWCNT & $-$ & $1.70\pm0.30$ & $61\pm7$ & $-$ & $100\pm30$ & $-$ & $-$ \\ \rowcolor{cyan!50}
    DS-IMDEA-MW\cite{Fernandez-Toribio2018} & Direct-spun few-layer MWCNT & $-$ & $1.10\pm0.20$ & $64\pm16$ & $-$ & $80\pm40$ & $-$ & $-$ \\ \rowcolor{cyan!50}
    DS-PUST-CSA\cite{Cho2018} & CSA 13\% stretched CNTF  & $-$ & $2.19\pm0.25$ & $176\pm22$ & $3.3$ & $39$ & $-$ & $-$ \\ \rowcolor{cyan!50}
    DS-PUST-Raw\cite{Cho2018} & Raw Direct-spun CNTF & $-$ & $0.95\pm0.13$ & $76\pm8$ & $5.5$ & $50$ & $-$ & $-$ \\ \rowcolor{cyan!50}
    DS-CAM-Den\cite{Koziol2007} & Acetone Direct-spun CNTF & $-$ & $1.1$ & $50$ & $3.8$ & $13$ & $-$ & $-$ \\ \rowcolor{cyan!50}
    DS-CAM-DogBone\cite{Motta2007a} & Dog Bone Direct-spun CNTF & $-$ & $2.2$ & $160$ & $-$ & $46$ & $-$ & $-$ \\ \rowcolor{cyan!50}
    VA-UT2004\cite{Zhang2004} & Twisted VA-CNTF & $0.8$ & $0.58$ & $--$ & $7.2$ & $14$ & $0.03$ & $-$ \\ \rowcolor{cyan!50}
    VA-UT2007\cite{Atkinson2007} & Twisted VA-CNTF & $0.8$ & $0.88$ & $--$ & $6$ & $27$ & $-$ & $-$ \\ \rowcolor{cyan!50}
    VA-MIT\cite{Hill2013} & Densified VA-CNTF & $0.8$ & $1.8$ & $89$ & $-$ & $94$ & $-$ & $-$ \\ \rowcolor{cyan!50}
    Graphene fiber\cite{Xin2015} & Optimized Graphene fiber & $1.87$ & $0.44\pm0.04$ & $--$ & $1.5$ & $-$ & $0.22\pm0.01$ & $1290\pm53$ \\ \rowcolor{cyan!50}
    Spider silk\cite{Vollrath2001} & Spider dragline silk & $1.3$ & $0.88\pm0.15$ & $131$ & $39.0\pm8.0$ & $165\pm30$ & $-$ & $-$ \\ \rowcolor{gray}
    Toray T300\cite{Mikhalchan2019} & PAN-based CF & $1.76$ & $2.01$ & $131$ & $1.5$ & $16$ & $0.06$ & $10.5$ \\ \rowcolor{gray}
    Toray T700SC & PAN-based CF & $1.8$ & $2.72$ & $128$ & $2.1$ & $-$ & $0.06$ & $9.6$ \\ \rowcolor{gray}
    Toray T800SC & PAN-based CF & $1.8$ & $3.27$ & $163$ & $2$ & $-$ & $0.08$ & $11.3$ \\ \rowcolor{gray}
    Toray T1000GB & PAN-based CF & $1.8$ & $3.54$ & $163$ & $2.2$ & $-$ & $0.07$ & $10.5$ \\ \rowcolor{gray}
    Toray T1100GC\cite{Mikhalchan2019} & PAN-based CF & $1.79$ & $3.91$ & $181$ & $2$ & $39$ & $0.07$ & $13$ \\ \rowcolor{gray}
    Toray M35JB\cite{Mikhalchan2019} & PAN-based CF & $1.75$ & $2.58$ & $196$ & $1.3$ & $18$ & $0.09$ & $38.9$ \\ \rowcolor{gray}
    Toray M40JB & PAN-based CF & $1.77$ & $2.49$ & $213$ & $1.2$ & $-$ & $0.1$ & $66.9$ \\ \rowcolor{gray}
    Toray M60JB\cite{Mikhalchan2019} & PAN-based CF & $1.93$ & $1.98$ & $305$ & $0.7$ & $7$ & $0.14$ & $150.5$ \\ \rowcolor{gray}
    HexTow AS4\cite{Mikhalchan2019,Wang2013} & PAN-based CF & $1.79$ & $2.48$ & $129$ & $1.7$ & $23$ & $0.06$ & $6.83$ \\ \rowcolor{gray}
    HexTow IM7 & PAN-based CF & $1.78$ & $3.2$ & $155$ & $1.8$ & $-$ & $0.07$ & $5.4$ \\ \rowcolor{gray}
    HexTow IM10 & PAN-based CF & $1.79$ & $3.81$ & $175$ & $2$ & $-$ & $0.08$ & $6.14$ \\ \rowcolor{gray}
    HexTow HM63 & PAN-based CF & $1.83$ & $2.64$ & $237$ & $1$ & $-$ & $0.11$ & $55$ \\ \rowcolor{gray!30}
    GRANOC YSH50A & Pitch-based CF & $2.1$ & $1.82$ & $248$ & $0.7$ & $-$ & $0.14$ & $120$ \\ \rowcolor{gray!30}
    GRANOC YSH70A & Pitch-based CF & $2.15$ & $1.69$ & $335$ & $0.5$ & $-$ & $0.2$ & $250$ \\ \rowcolor{gray!30}
    GRANOC YS80A & Pitch-based CF & $2.17$ & $1.67$ & $362$ & $0.5$ & $-$ & $0.2$ & $320$ \\ \rowcolor{gray!30}
    GRANOC YS90A & Pitch-based CF & $2.18$ & $1.62$ & $404$ & $0.3$ & $-$ & $0.33$ & $500$ \\ \rowcolor{gray!30}
    GRANOC YS95A & Pitch-based CF & $2.19$ & $1.61$ & $420$ & $0.3$ & $-$ & $0.45$ & $600$ \\ \rowcolor{gray!30}
    DIALEAD K1352U & Pitch-based CF & $2.12$ & $1.7$ & $292$ & $0.6$ & $-$ & $0.15$ & $140$ \\ \rowcolor{gray!30}
    DIALEAD K1392U & Pitch-based CF & $2.15$ & $1.72$ & $353$ & $0.5$ & $-$ & $0.2$ & $210$ \\ \rowcolor{gray!30}
    DIALEAD K13C6U & Pitch-based CF & $2.18$ & $1.65$ & $413$ & $0.4$ & $-$ & $0.4$ & $580$ \\ \rowcolor{gray!30}
    DIALEAD K13D2U & Pitch-based CF & $2.2$ & $1.68$ & $425$ & $0.4$ & $-$ & $0.67$ & $800$ \\ \rowcolor{orange!50}
    Zylon AS\cite{Wang2013} & PBO fiber & $1.54$ & $3.7$ & $117$ & $3.5$ & $65$ & $-$ & $19$ \\ \rowcolor{orange!50}
    Zylon HM\cite{Wang2013} & PBO fiber & $1.56$ & $3.7$ & $172$ & $2.5$ & $46$ & $-$ & $23$ \\ \rowcolor{teal!70}
    Dyneema SK60 & UHMWPE fiber & $0.97$ & $3.1$ & $107$ & $3.7$ & $-$ & $-$ & $20$ \\ \rowcolor{teal!70}
    Dyneema SK75\cite{Mikhalchan2019,Wang2013} & UHMWPE fiber & $0.97$ & $3.74$ & $133$ & $3.5$ & $63$ & $-$ & $14$ \\ \rowcolor{teal!70}
    Dyneema SK99 & UHMWPE fiber & $0.98$ & $4.3$ & $159$ & $3.5$ & $-$ & $-$ & $20$ \\ \rowcolor{magenta!70}
    Vectran UM\cite{Mikhalchan2019} & Polyarylate fiber & $1.41$ & $2.03$ & $73$ & $2.7$ & $30$ & $-$ & $-$ \\ \rowcolor{magenta!70}
    Vectran HT\cite{Mikhalchan2019} & Polyarylate fiber & $1.41$ & $2.29$ & $53$ & $3.8$ & $43$ & $-$ & $1.5$ \\ \rowcolor{olive!70}
    Kevlar 29 & Polyamide fiber & $1.44$ & $2.03$ & $49$ & $3.6$ & $36$ & $-$ & $0.04$ \\ \rowcolor{olive!70}
    Kevlar 49 & Polyamide fiber & $1.44$ & $2.08$ & $78$ & $2.4$ & $25$ & $-$ & $0.04$ \\ \rowcolor{olive!70}
    Kevlar 149\cite{Hearle2008} & Polyamide fiber & $1.47$ & $2.35$ & $122$ & $2.5$ & $-$ & $-$ & $3$ \\ \rowcolor{olive!70}
    Technora\cite{Mikhalchan2019,Wang2013} & Polyamide fiber & $1.39$ & $2.4$ & $54$ & $4.2$ & $42$ & $-$ & $-$ \\ \rowcolor{pink}
    C-Glass & Glass fiber & $2.54$ & $1.3$ & $27$ & $4.8$ & $1.1$ & $-$ & $-$ \\ \rowcolor{pink}
    E-Glass\cite{Mikhalchan2019} & Glass fiber & $2.57$ & $1.41$ & $28$ & $4.8$ & $1.3$ & $-$ & $-$ \\ \rowcolor{pink}
    S-Glass\cite{Mikhalchan2019} & Glass fiber & $2.48$ & $1.85$ & $35$ & $5.4$ & $1.45$ & $-$ & $31$ \\ \rowcolor{pink}
    R-Glass & Glass fiber & $2.54$ & $1.63$ & $34$ & $4.8$ & $-$ & $-$ & $37$ \\ \rowcolor{gray!70}
    Gold\cite{Behabtu2013} & Metal & $19.32$ & $0.01$ & $4$ & $-$ & $-$ & $45.3$ & $315$ \\ \rowcolor{gray!70}
    Silver\cite{Behabtu2013} & Metal & $10.5$ & $0.01$ & $7$ & $-$ & $-$ & $62$ & $427$ \\ \rowcolor{gray!70}
    Copper\cite{Behabtu2013} & Metal & $8.92$ & $0.04$ & $13$ & $-$ & $-$ & $59.4$ & $398$ \\ \rowcolor{gray!70}
    Nickel\cite{Behabtu2013} & Metal & $8.73$ & $0.04$ & $19$ & $-$ & $-$ & $15.36$ & $91.4$ \\ \rowcolor{gray!70}
    Aluminium\cite{Behabtu2013} & Metal & $2.72$ & $0.04$ & $25$ & $-$ & $-$ & $47.6$ & $238$ \\ \rowcolor{gray!70}
    Stainless steel & Metal & $7.8$ & $0.35$ & $29$ & $1.4$ & $-$ & $1.5$ & $17$ \\
    \botrule
    \end{tabular*}
    \tiny
    \footnotetext[1]{Most data of commercial fibers are based on the corresponding technical data from their company's websites or www.matweb.com}
    \end{minipage}
    \end{center}
    \end{sidewaystable}

\section*{S4. Experiment details on the In-Situ Stretching Raman.}\label{S4}
Monitoring the redshift of Raman mode is often the clearest and simplest characterization to detect strain on iCNTs.
\par
During the In-Situ Stretching Raman (ISSR) characterization, the suspended CNTFs are ends fixed onto a manual stretching stage to detect the Raman signal with HORIBA HR800 micro-Raman spectroscopy. We excite the Raman G' mode with linearly polarized laser with power on sample $\rm \sim 4.9\, mW$. Because the redshift level is proportional to the strain. Among the Raman modes, the two phonon processes, such as the G' mode, possess a higher redshift rate ($\rm \sim 70 \,cm^{-1}\%^{-1}$), which increases the accuracy of the strain distribution measured. Moreover, the insensitivity of G' mode on the types of CNTs also facilities the deducing of strain.
\par
Within CNTFs, the diameter of CNTs is $\rm \sim 10\, nm$. The laser spot size ($\rm \sim 2\, \mu m$) covers thousands of tubes and length in micron scale in each tube. Therefore, the Raman scattering signal gives the strain distribution among these thousands of tubes and along each iCNTs. Because strain, which is proportion to the content of redshift, is unevenly distributed inter and intra tubes, the redshift signal detected gives the strain distribution normalized by volume. Because the fiber is isotropic across the depth (as verified by cross-section cut by FIB), the results of ISSR offers the overall description of the fiber. Moreover, for iCNTs, the linewidth of Raman mode will not increase even with a strain at least $4\%$,\cite{Chang2010} thus, the broadening of Raman peak with stretching can be converted to the broadening of strain distribution instead of the linewidth broadening of Raman mode itself. Therefore, we can use the collective signal to analyze the distribution of strain among CNTs.
\par
\begin{figure}[H]%
    \centering
    \includegraphics[width=0.9\textwidth]{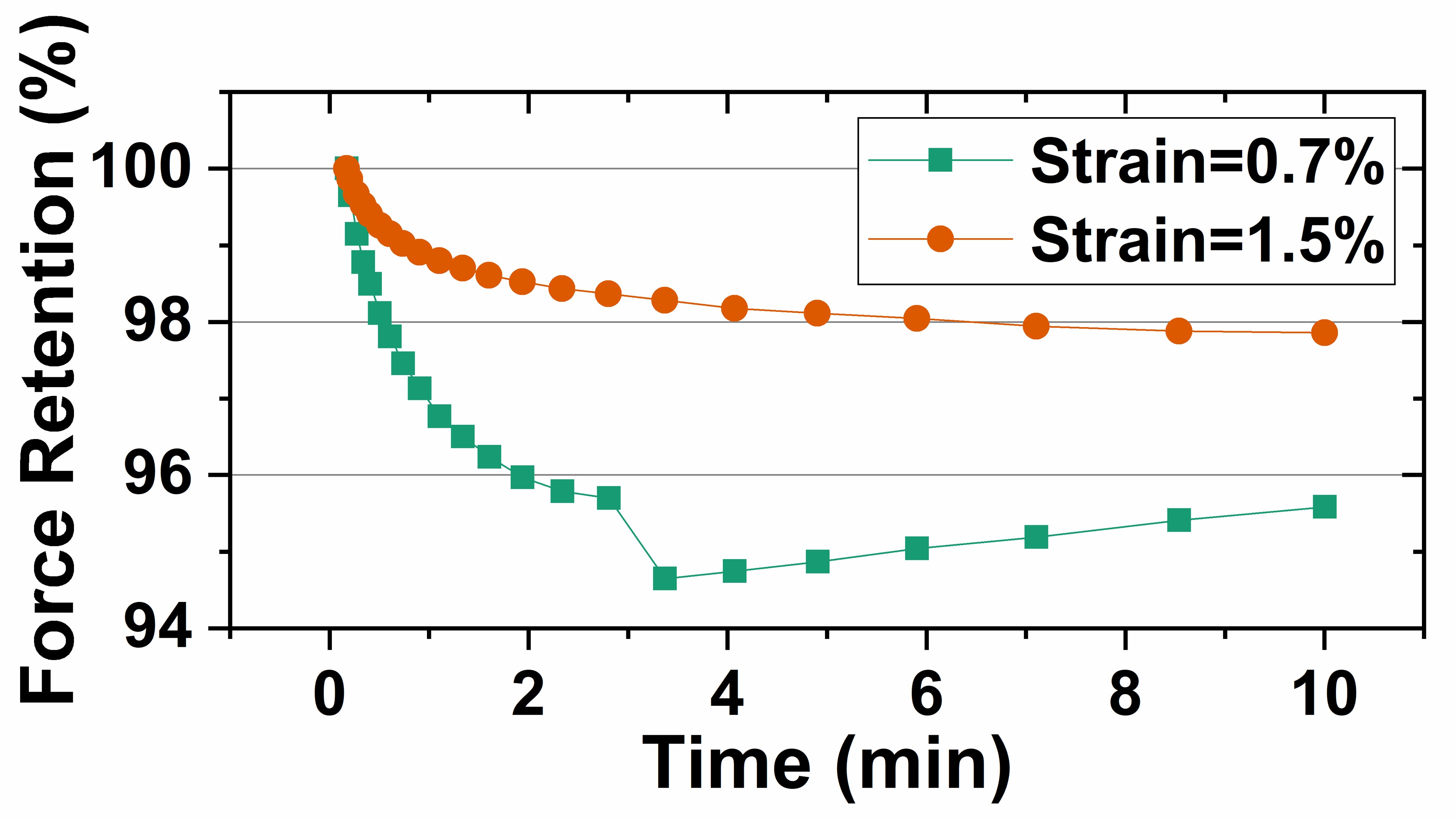}
    \caption{\textbf{Force retention of raw CNTF under tension.} The strain of $0.7\%$ and $1.5\%$ is exerted onto raw CNTF to monitor the change of force. As shown, the force can be maintained to at least $95\%$ within 10 mins.}\label{Fig:S8}
\end{figure}

Here, with a Glan Polarizer, we only collect the scattered radiation in the parallel polarization with the laser, so that only iCNTs with their axis nearly along the laser polarization can be detected. For the ZZ/XX configuration,\cite{Damnjanovic1999} both incident and scattered photons detected parallel/perpendicular to the axis of CNTFs, offers strain distribution of CNTs along/normal to the fiber axis. Additionally, we have not noticed the relaxation of strain during the ISSR experiment. At least $95\%$ of force retention within 10 mins is verified by the strain relaxation experiment (Fig.S\ref{Fig:S8}). The force retention is sufficient for us to complete the ISSR characterization.
\par

\section*{S5. Details on the enhancing mechanism.}\label{S5}
\subsection*{S5.1: Why the raw CNTFs are weak on mechanical properties:}\label{S5.1}
When a raw CNTF is under load, tensioning lines frequently appear from the disordered iCNTs network in raw CNTFs, which indicates the stress concentration. The tensioning lines are also the small load-bearing portion of CNTs as indicated by ISSR (Fig.7c). Around the tensioning line, the slippage and failure happen first. This is because the tensioning line, as the shortest pathway, is the first region of the material to take the load and alone.
\par
In the fibrillar structure of CNTFs, the rupture procedure under increasing load proceeds with the shortest bundle breaking first, then the second-rated shortest bundle takes the increased load, and so forth. Eventually, as the major distribution of bundle lengths is approached and broken, there will be insufficient bundle left to support the load, and catastrophic failure will occur. Because in raw CNTFs, at any time of stretching, only a small portion of bundles are shortest within the disordered network. During the whole stretching process, the bundles just successively break with a unit of small portion of them. Therefore, the tangent modulus at any time of stretching is low.
\par

\subsection*{S5.2: $\rm 1^{st}$ Enhancing mechanism: the increased proportion of load-bearing bundles.}\label{S5.2}
As shown within a simplified CNTF cell (Fig.8a), if the load is exerted on the vertical surfaces of the cell (along axis), because there is no medium between CNTs to transmit the load, only the shortest CNT bundle is loaded (bundle \textcircled{3}, the red lines). The curved bundles (bundle \textcircled{2}\textcircled{3}\textcircled{4}) do not substantially contribute to bearing the load, despite portions of them being oriented in the vertical direction (circled by blue dots line). As observed in the ISSR, a substantial proportion of CNTs do not share the load, particularly for the raw CNTF with disordered microstructure.
\par
After the drawing in CSA, more crumpled tubes and bundles are straightened and become effective to link the “shortest” distance (the red lines in Fig.8b). They jointly participate in sharing the load after the initial stretching, indicated by the soaring of initial modulus (Fig.4b, d). This optimization is also verified by the entire redshift of G' mode in Zone i' (Fig.7e and g) and the larger portion of tubes further redshift in Zone ii'. As the load continues to increase, effective bundles fail successively, decreasing the modulus monotonically (Fig.4b). The failed bundles return to the initial state free of strain, as observed in the redshift release in Zone ii' (Fig.7b and e blue dotted arrow).
\par

\subsection*{S5.3: $\rm 2^{nd}$ Enhancing mechanism: the extension of effective length of tubes in load-bearing bundles.}\label{S5.3}
Only accounting for the increased fraction of load-bearing bundles in network does not fully explain the much higher final/maximum strain on iCNTs, i.e., $\varepsilon\rm_i^*$ for the DD-CNTF. Another factor must be considered for the mechanism, particularly considering the maximum strain turns up just before the failure. The much-improved tenacity value for the fully DD-CNTF must link to this factor.
\par
We attribute the optimization of $\varepsilon\rm_i^*$ to the effective CNTs length increase within load-bearing bundles, which results in a higher activation barrier for slippage. As found recently by Yakobson et al, the product of the tube length and the friction coefficient between tubes is positively correlated with the tensile properties of a bundle.\cite{Gupta2020} For the elastic interface (static friction), the stress exerted on iCNTs $\sigma _i=Y \varepsilon\rm_i$, is balanced by the friction from the surrounding tubes, where $Y$ and $\varepsilon\rm_i$ is the tube's Young's modulus and strain, respectively. When $\sigma \rm_i$ increases to the critical value $\sigma\rm _i^*$, slippage will occur and the elastic interface will begin to deform plasticly. Immediately prior to slippage, $\sigma{\rm_i^*}\,A{\rm_t}=f{\rm_s}\,L{\rm_{eff}}$, where At is the cross-sectional area of the tube, fs is tube's maximum static friction coefficient per unit length, and $L{\rm_{eff}}$ is the effective length of tube that sharing the load (friction). It is easy to find that $\sigma\rm _i^*$ is the maximum value for $\sigma \rm_i$, because $L{\rm_{eff}}$ will shrink as the tube-tube contact area drops during relative slipping. Therefore, the corresponding $\varepsilon\rm_i^*$ can be deduced by:
\par

\begin{equation}\label{eq:S3}
    \varepsilon{\rm_i^*}=\frac{\sigma{\rm_i^*}}{Y}=\frac{f{\rm_s}\,L{\rm_{eff}}}{Y\,A{\rm_t}}
    \tag{S3}
\end{equation}

\par
Previous models often contain an implied assumption that the entire lengths of tubes ($L$) are attached to the bundle, i.e., $L{\rm_{eff}}=L$. However, in a fibrillar structure, particularly raw CNTF, tubes within the hierarchical network only partly align with any specific bundle, and may be incorporated into many bundles. Particularly, for the CNTF used in this work, tubes produced by FCCVD have long lengths, $\sim100\,\mu m$.\cite{Mikhalchan2019} Because the load can only be transmitted through the coupling between adjacent tubes, only the tube section attached to an effective bundle can participate in sharing the load (the red lines as indicated by solid blue arrows in Fig.8), i.e., $L{\rm_{eff}}$ is always smaller than $L$. In contrast, the dangling CNTs or fraction of CNTs attached to an ineffective bundle (hollow blue arrow) do not transmit load, even though these portions may belong to the same tube. As observed in Fig.1b-c, in most cases for raw CNTF, $L{\rm_{eff}}\gg L$.
\par
With the Double-Drawing process, the crumpled tubes are extended by drawing and squeezed together by Poisson Tightening, forming bundles with large diameter (red aggregated bundle [\textcircled{2}\textcircled{3}\textcircled{4}] in Fig.8b, Fig.1d-e and Fig.2a). Thus, longer length of tubes aggregates into the effective bundles, with $L{\rm_{eff}}$ approaching $L$. As per Equation (\ref{eq:S3}), an increase in $L{\rm_{eff}}$ proportionally increases $\varepsilon\rm_i$ needed to activate the slippage, which delays the failure of DD-CNTF and improves the tenacity.
\par

\bibliography{library}